%% file: jcap_ver.tex
\documentclass[a4paper,11pt]{article}
\pdfoutput=1

\usepackage{jcappub} 
\usepackage[T1]{fontenc} 
\usepackage{float}
\usepackage{lineno}

\title{\boldmath Search for the imprint of axion-like particles in the highest-energy photons of hard $\gamma$-ray blazars}

\author[a]{R. Buehler}
\author[a]{G. Gallardo}
\author[a]{G. Maier}
\author[b]{A.Dom\'inguez}
\author[b]{M. L\'opez}
\author[c]{M. Meyer}

\affiliation[a]{Deutsches Elektronen-Synchrotron, D-15738 Zeuthen, Germany}
\affiliation[b]{IPARCOS and Department of EMFTEL, Universidad Complutense de Madrid, E-28040 Madrid, Spain}
\affiliation[c]{Friedrich-Alexander-Universit\"{a}t Erlangen-N\"{u}rnberg, Erlangen Centre for Astroparticle Physics,Erwin-Rommel-Str. 1, 91058 Erlangen, Germany}

\emailAdd{rolf.buehler@desy.de}
\emailAdd{iridiumraven@gmail.com}
\emailAdd{gernot.maier@desy.de}
\emailAdd{alberto.d@ucm.es}
\emailAdd{marcolop@ucm.es}
\emailAdd{manuel.e.meyer@fau.de }

\abstract{Axion-like particles (ALPs), predicted in theories beyond the Standard Model, can have observational effects on the transparency of the Universe to $\gamma$ rays in the presence of magnetic fields. In this work, we search for effects compatible with the existence of ALPs with 80 months of data from the {\it Fermi} Large Area Telescope, by comparing the distributions of observed highest energy photons from sources beyond redshifts of z $\geq$ 0.1 with theoretical predictions in the presence of ALPs. We find no evidence for an increased $\gamma$-ray transparency due to ALPs and therefore we set limits on the ALPs parameters assuming a value of the intergalactic magnetic field strength of 1 nG. Photon-ALP couplings above $10^{-11}$ GeV$^{-1}$ are excluded for ALP masses $m_{a}$ $\lesssim 3.0$ neV. As the allowed magnetic field parameter space is large, we also test lower magnetic field strengths and no constraints can be set for B$\leq$0.1 nG below the CAST limit. These constraints exclude a region of the parameter space not covered by other $\gamma$-ray telescopes and are compatible with limits imposed by other experiments.}


\begin{document}
\maketitle
\flushbottom

\section{\label{sec1}Introduction}

Axion-like particles (ALPs), very light pseudo-scalar bosons predicted by multiple extensions of the Standard Model \cite{ALPS:btsm1,ALPS:btsm2,ALPS:btsm3,ALPS:btsm4,ALPS:btsm5,ALPS:btsm6}, could be detected through their coupling to photons in the presence of external magnetic fields, 
\begin{equation} 
\mathcal{L}_{a\gamma}=-\frac{1}{4}g_{a\gamma}F_{\mu\nu}\tilde{F}^{\mu\nu}a=g_{a\gamma}\mathbf{E}\cdot\mathbf{B}a,\label{eq1}
\end{equation}
where $g_{a\gamma}$ is the axion-photon coupling strength, $F_{\mu\nu}$ is the electromagnetic tensor, $\mathbf{E}$ and $\mathbf{B}$ are the electric and magnetic fields, and $a$ is the axion-like particle field.

As their name suggests, these particles are a generalization of the quantum chromodynamics (QCD) axion, predicted by the Peccei-Quinn mechanism in order to solve the strong CP problem \cite{AX:PQ,cor7,cor8}. In contrast to axions, the mass and the coupling constant of ALPs are completely independent parameters \cite{2014arXiv1407.0546R;AXIONS_OVERVIEW_RINGWALD}. ALPs are also a cold dark matter candidate \cite{PRESKILL1983127,ALPS:btsm6,11n2:dine,11n3:abbott,Sikivie:2009fv} for certain values of the mass and the coupling.

Due to their coupling in Eq. \ref{eq1}, ALPs can affect the propagation of $\gamma$-ray photons coming from astrophysical sources. Within
the Standard Model, these photons are absorbed by pair production processes with the extragalactic background light (EBL). This interaction causes an attenuation of the spectra of $\gamma$-ray sources that increases with the energy of the $\gamma$ rays and the distance to the source. As a consequence, the transparency of the Universe to $\gamma$ rays decreases \cite{EBL:bitteau,EBL:dwek}. This transparency is quantified in the cosmic $\gamma$-ray horizon, defined as the isocontour in the energy-redshift plane where the optical depth equals 1. Beyond this line, we do not expect many surviving photons according to conventional EBL models.

Once produced, $\gamma$  rays could oscillate into ALPs in different astrophysical magnetic field environments  that have been observed by several experiments \cite{FIELDS:faraday,IGM:ns_cmb4,IGM:ns_or1,PAG:IC_obs,bfields}. Such fields include the region surrounding the emitting $\gamma$-ray source, the medium within a galaxy cluster, the intergalactic medium and the Milky Way \cite{FIELDS:rev}. When a photon turns into an ALP, an ALP is not affected by the EBL and thus it can travel cosmological distances unhindered. It may then oscillate back into a photon, leading to a modification of the transparency of the Universe to $\gamma$ rays. It is useful to define a critical energy, given by \cite{TheFermi-LAT:2016zue,hooper2007},
\begin{equation}
E_{c}(\mathrm{GeV})\sim2.5\frac{|m_{a,\mathrm{neV}}^{2}-\omega_{\mathrm{\mathit{pl},neV}}^{2}|}{2g_{11}B_{T,\mathrm{\mu G}}},\label{eq4}
\end{equation}
with ALP mass $m_{a,\mathrm{neV}}=m_{a}/\mathrm{neV}$, rescaled coupling constant $g_{11}=g_{a\gamma}/10^{-11}\,\mathrm{GeV^{-1}}$and transversal field component $B_{T,\mathrm{\mu G}}=B_{T}/\mathrm{\mu G}$. The second term in the numerator, $\omega_{pl}=\sqrt{4\pi\alpha n_{e}/m_{e}}$, is the plasma frequency of the medium\footnote{The photon obtains an effective mass while propagating through the cold plasma of electrons.} in units of $\mathrm{neV}$, where $n_{e}$ is the electron density, $m_{e}$ is the electron mass and $\alpha$ denotes the fine structure constant. Above $E_{c}$, the photon-ALP conversion probability becomes maximal, causing a hardening of the spectra of $\gamma$-ray sources \cite{hardening1,hardening2,hardening3,paper_igmf,hardening4}. Below $E_{c}$, the mixing can also induce spectral irregularities for certain magnetic field scenarios \cite{TheFermi-LAT:2016zue,hess_limits}.

In this work, we study the transparency of the Universe to $\gamma$ rays making use of the highest-energy photon (HEP) events measured with the Large Area Telescope (LAT) on board the {\it Fermi} satellite \cite{2009ApJ...697.1071A:fermilat_atwood2009}. For each source and over a time period, the \emph{Fermi}-LAT measures a set of photons with energies $\left\{ E_{1},E_{2},...,E_{N}\right\} $, from which the HEP is the event with maximum energy: $\max\left\{ E_{1},E_{2},...,E_{N}\right\} $. Through a maximum likelihood analysis, we compare the simulated HEP distributions in the presence and absence of ALPs. We assume that the conversion takes place in the intergalactic magnetic field (IGMF), which yields critical energies from $\sim100$ GeV up to the TeV range for values of the coupling constant above $g_{11}\sim1$ and masses below $m_{a,\mathrm{neV}}\sim3$. Conversions in our galaxy is not taken into account since the mixing only affects the spectra of blazars at TeV energies for $m_{a}\geq1$ neV. Photon-ALPs interactions within the jet of the blazar can lead to a reduction in photon flux, which means an additional initial state of ALPs in the beam that could oscillate into photons in the IGMF region. The estimated conversion probability for BL Lacs has a complex dependency on the magnetic field, source and ALPs parameters, whereas it is negligible at high energies for FSRQs \cite{PAG:agn}. Due to this and the lack of information regarding all the sources in our sample, we do not consider mixing in this region. However, depending on the source and the ALPs parameters, the mixing effects could modify our results by $10$\%$-30$\%.

This paper is organized as follows. Section \ref{sec2} outlines the physical interactions relevant for the propagation of $\gamma$-ray photons in the intergalactic medium. Section \ref{sec3} contains a description of the data and the simulations procedure. In Section \ref{sec4}, we give a detailed explanation of the likelihood analysis. The results obtained in our analysis are presented and discussed in Section \ref{sec5}.

\section{\label{sec2}Mixing in the intergalactic medium}

\subsection{Photon annihilation with the extragalactic background light}

The EBL is the accumulated radiation in the Universe from the infrared to the ultraviolet wavelengths. This background radiation has its roots in stars formation processes, AGN and the starlight re-processed by dust in galaxies \cite{EBL:dwek,EBL:bitteau,2017A&A...603A..34F:EBL_FRV_current}.

The flux of extragalactic $\gamma$-ray sources is attenuated due to electron-positron pair production processes that occur because of the interaction of $\gamma$-rays with EBL photons. \cite{pair_production,cor9}. Because of the cross section of pair production and the wavelength range of the EBL, the latter plays a key role in the observation of the $\gamma$-ray sky, being the fundamental source of opacity for the Universe to $\gamma$ rays. It is also relevant for re-ionization models in cosmology and galaxy formation and evolution \cite{EBL:dwek,cor3,EBL:sfh2,pubv2}.

Direct measurements of the EBL are challenging due to the presence of other backgrounds in the solar system and our galaxy, such as the bright zodiacal light and the galactic emission \cite{EBL:challenges}. The main constraints come from the integrated light of discrete extragalactic sources and from $\gamma$-ray observations \cite{EBL:latest,EBL:magic1,EBL:veritas1,EBL:lat1}, for which blazars are good probes \cite{cor6}. Many different approaches have been used to model the intensity and spectral shape of the EBL. In this work, we use the observationally-based Dom\'inguez et al. model \cite{EBL:dominguez} to derive all the results and the Finke et al. \cite{EBL:forward1} model for comparison. 

The survival probability, or attenuation factor, is described by a decreasing exponential law, $P_{\gamma\gamma}=\exp\left[-\tau(E,z)\right]$. It depends on the optical depth parameter $\tau(E,z)$, which is an increasing function of the photon energy and the distance to the source. The $\gamma$-ray spectra of sources are then described by, 
\begin{equation} 
\phi_{obs}(E)=\phi_{int}(E)\cdot\exp\left[-\tau(E,z)\right],\label{eq2}
\end{equation}
where $\phi_{obs}$ and $\phi_{int}$ are the observed and intrinsc spectra, respectively. The latter would be the observed spectrum of the source if there was no EBL absorption. The cosmic $\gamma$-ray horizon (CGRH) \cite{cor4,cor5} is defined as the energy $E_{0}$ at which the optical depth becomes unity $\tau(E_{0},z)=1$. The solid line in Fig. 13 of Ref. \cite{2FHL:catalog} is the CGRH derived with the Dom\'inguez et al. model. Above the CGRH curve, the Universe is more opaque to $\gamma$ rays due to larger values of $\tau$ that translate into smaller photon survival probabilities, whereas in the region below the curve, the survival probabilities are larger and the Universe is more transparent to $\gamma$ rays. For a given redshift, the CGRH quantifies the maximum energy of photons that survive the EBL. If there were modifications of the canonical $\gamma$-ray propagation, the observed HEP event for each source should change correspondingly. This is what we use in order to search for ALPs effects. 

\subsection{\label{sec2.2}Photon-ALP mixing} 
The photon-ALP system is described by the lagrangian of Eq. \ref{eq1}. Given a homogeneous magnetic field $\mathbf{B}$ over a distance of length $s$ and a polarized photon beam, the photon-axion conversion probability $P_{\gamma\rightarrow a}$ can be calculated analytically \cite{ALPS:oscillations},
\begin{equation}
P_{\gamma\rightarrow a}=\frac{1}{1+\left(\frac{E_{c}}{E}\right)^{2}}\sin^{2}\left(\frac{g_{\gamma a}B_{T}s}{2}\sqrt{1+\left(\frac{E_{c}}{E}\right)^{2}}\right),\label{eq3}
\end{equation}
where $B_{T}$ is the transverse component to the direction of propagation, $g_{\gamma a}$ is the coupling constant, $E$ is the photon energy and $E_{c}$ is the critical energy around which we expect the ALPs effects to occur, defined in Eq. \ref{eq4}.

Equation \ref{eq3} is only valid if the photon-ALP beam is in a pure polarization state and the conversion takes place in a homogeneous magnetic field. Since photon polarization is not measured in the $\gamma$-ray band, we have to treat the photon-ALP beam as unpolarized and take into account general magnetic field morphologies. 

The IGMF is often modeled as a domain-like structure in redshift, with a homogeneous magnetic field strength in each cell and a random orientation in different domains \cite{paper_igmf}. The size of each cell is the so-called coherence length, a distance upon which the magnetic field is homogeneous. Unfortunately, there are only upper limits available for the strength and coherence length of large-scale magnetic fields, with $B\leq1$ nG and $s\sim$ Mpc \cite{IGM:ns,IGM:ns_or1}. These upper limits are used as model parameters for the analysis performed in this work, $B=1$ nG and $s=1$ Mpc. The analysis is also carried out with different values of the magnetic field parameters in order to evaluate its systematic uncertainties. We take $n_{e}\sim10^{-7}\:\mathrm{cm^{-3}}$ as the electron density in the intergalactic medium \cite{IGM:cosmo_book}. With this value of $n_{e}$, the plasma frequency is $\omega_{pl}\sim1.17\cdot10^{-14}$ eV.

The propagation of the beam in a domain-like structure is a stochastic process. For this reason, the effects of a single trajectory of the beam cannot be measured. An average over a large number of realizations of the process is required in order to evaluate the photon survival probability. We compute the photon-ALP oscillation probability in the transfer-matrix formalism as described in, e.g., Refs. \cite{GOLD:formulas,cor10,PAG:IC_manuel}, using the \emph{$\mathsf{gammaALPs}$} code\footnote{\url{https://github.com/me-manu/gammaALPs}}. A summary is provided in Appendix A. The systematic uncertainties associated with the magnetic field parameters $B_{T}$ and $s$ have the biggest impact on the results, changing the exclusion region size up to ~35\%, as discussed in Section \ref{sec5} and in Appendix B. Additionally, instead of using a single magnetic field realization for all the sources, we use different realizations for the different lines of sight.

\section{\label{sec3}Data and simulations}

We use the energy of the HEP events from the Second Catalog of Hard \footnote{Sources with spectral index smaller than 2. Hard sources are easier to observe at higher energies.} {\it Fermi}-LAT sources (2FHL). The catalog reports the properties of 360 sources significantly detected by the LAT above 50 GeV from August 2008 to April 2015 \cite{2FHL:catalog}. The characterization of such sources at these energies was made possible with the Pass 8 event-level analysis \cite{LAT:pass8} and long telescope exposure. More than 80\% of the sources are extragalactic, of which 75\% are AGN.

We chose this set of sources because the critical energy of the ALPs mixing in the IGMF lies within the energy range of the 2FHL catalog, for the IGMF field parameters, couplings $g_{11}$ and masses $m_{a,\mathrm{neV}}$ tested in this work. Out of these available sources we only use source  with redshifts $z\geq0.1$. For smaller values of redshift the effects of ALPs are too small to be detected. The catalog also offers the probabilities of each photon event belonging to each one of the sources in the region of interest. We only take sources for which the probability of the HEP assigned to them is $P\geq0.99$. This cut improves the background rejection but it also entails a reduction in statistics, from a sample of 96 sources to 79 sources. With this cut, results are more conservative and the excluded region decreases by ~15\%, see Appendix \ref{AB} for more details. 

An overview of the method goes as follows. We require two main elements, namely, the HEP events from the 2FHL catalog sources and their corresponding probability distribution functions (p.d.f.s). The former are taken directly from the catalog results, whereas the latter are derived from simulations. Then, we compute the probability or likelihood that each HEP has with its corresponding distribution, and we combine them because the sources are independent.

\begin{figure}[H]
\centering 
\includegraphics[width=1\textwidth]{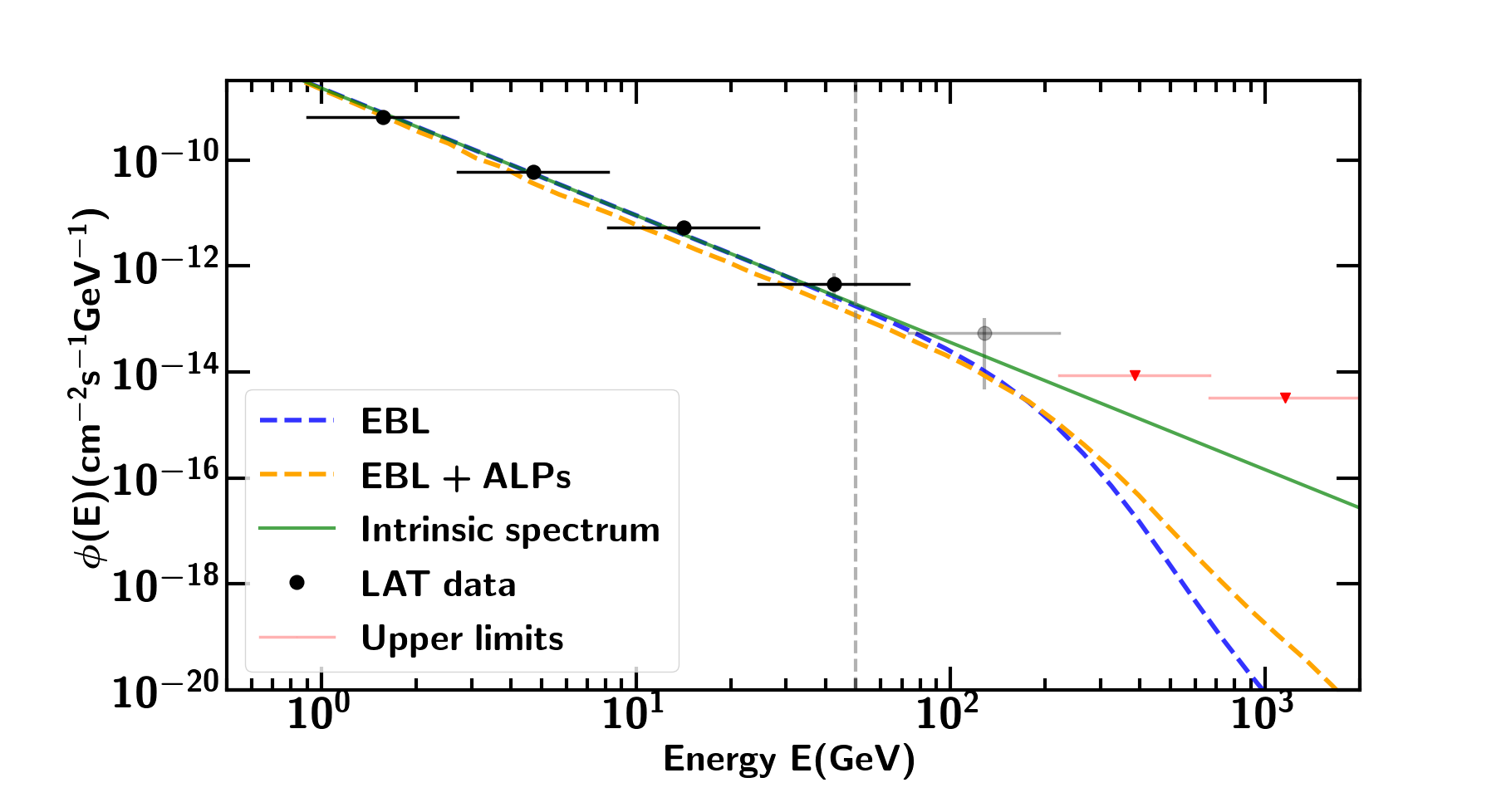}
\caption[Results]{Energy spectrum for 2FHL 2000.9-1749, source  located at $z=0.65$. The  points (black are fitted, grey are excluded and red are upper limits) are the {\it Fermi}-LAT data from Ref. \cite{BLZ:spectra} and their corresponding spectral fit. The dashed lines represent the absorbed spectrum in the absence and in the presence of ALPs with $m=1$ neV and $g_{11}=7$. The EBL model used for the attenuation is the Dom\'inguez et al. model. The dashed gray line at 50 GeV represents the energy at which we start for this analysis. Below $E_c$, the effect of ALPs is an extra dimming of the source, this is why the yellow line is beneath the others in this energy range. Fitting the EBL+ALPS spectrum to the low energy data points would result in a somewhat larger number of expected high energy photons and hence stronger constraints. We have not performed this rescaling, which requires fitting the Fermi-LAT spectra with the ALP model.}     
\label{fig:spr}  
\end{figure}

For each source, we simulate a HEP probability distribution from which we expect the measured HEP to come. We use 40 logarithmically spaced energy bins from 50 GeV to 2 TeV, the same range as in the catalog, and we compute the expected number of events between energies $E_{1}$ and $E_{2}$ through 
\begin{equation}
N_{E_{1}E_{2}}=\int_{E_{1}}^{E_{2}}P_{\gamma\gamma}\left(E,z,\theta\right)\phi(E)\epsilon(E)dE.\label{eq5}
\end{equation}
The first term in the equation, $P_{\gamma\gamma}\left(E,z,\theta\right)$, is the photon survival probability from Eq. \ref{eq2}. The parameter $\theta=(m_{a},g_{a\gamma})$ represents the mass of the axion and the coupling constant for the ALPs model of Section \ref{sec2.2}. The second term, $\phi(E)$, is the intrinsic spectrum of the source, which has to be derived in a region with negligible EBL attenuation, below the range of the catalog. Therefore, we make use of a recent re-analysis of the 2FHL sources, carried out in Ref. \cite{BLZ:spectra}, which extends the analysis spectral range down to 300 MeV. Following a $\chi^{2}$ minimization procedure, we fit these spectral data to power laws and logarithmic parabolas. This is possible, as $\gamma$-ray sources typically have smooth spectra over a limited energy range and blazars are well described by these functions \cite{EBL:sfh2,EBL:lat1}. The fits take spectral points until $\tau(z,E)\sim0.1$, energies in which EBL effects cease to be negligible. A list of the sources and the spectral fits can be found in Appendix C. An example of the spectral fit and absorption models on the spectrum of a source is displayed in Fig. \ref{fig:spr}.

\begin{figure}[H]
\centering 
\includegraphics[width=0.8\textwidth]{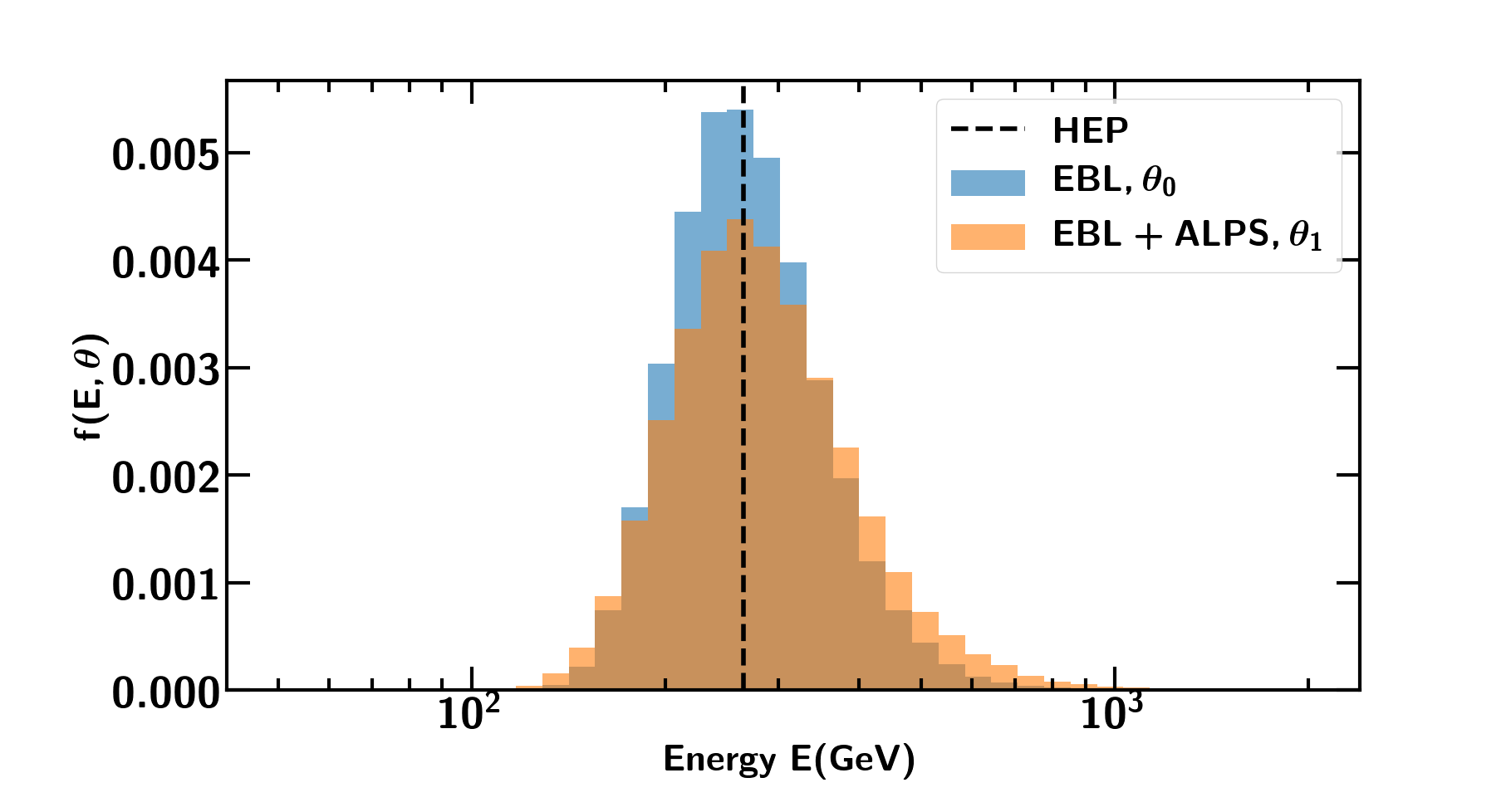} 
\caption[Results]{Simulated HEP p.d.f.s for 2FHL J0222.6+4301, located at $z=0.444$. The blue histogram represents the null hypothesis, whereas the orange histogram includes ALPs with $m=1$ neV and $g_{11}=7$.  The dashed black line is the observed HEP by the  {\it Fermi}-LAT, from Ref. \cite{2FHL:catalog}.}     
\label{fig:spdf}  
\end{figure}

The last term in Eq. \ref{eq5}, $\epsilon(E)$, is the exposure map of the {\it Fermi}-LAT, an integral of the total instrument response function over the entire region of interest of each source, provided in the analysis files of Ref. \cite{2FHL:catalog}. The probability of detecting $c$ counts in the $i$-th bin is given by Poisson statistics, i.e. $p_{ij}=N_{ij}^{c}\exp(-N_{ij})/c!$, where the index $j$ stands for the source and $N_{ij}$ is given by Eq. \ref{eq5}. Using these probability functions, we generate events for each source and energy bin. The last non-empty energy bin, $c\geq1$, is taken as the bin with the HEP. For each source $j$ and attenuation model $\theta$, we build a histogram of HEPs by repeating these pseudo-experiments $10^{4}$ times. These normalized histograms are the HEP p.d.f.s. An example is displayed in Fig. \ref{fig:spdf}.

\section{\label{sec4}Likelihood analysis}

For a random variable $x$ distributed according to a p.d.f. $f(x,\theta)$, where $\theta$ is any parameter of the function, the probability for a measurement $x_{i}$ to be in {[}$x_{i},x_{i}+dx_{i}${]} is $f(x_{i},\theta)dx_{i}$. Assuming $N$ independent observations of $x$, the joint likelihood function is \cite{cowan:stats}, 
\begin{equation}
L(x_{1},x_{2},...,x_{N}|\theta)=\prod_{j=1}^{N}f_{j}(x_{j},\theta).
\end{equation}

In our work, the random variable is the HEP of each source, $E_{j}$, which is the maximum energy event with a probability larger than 99\% to be associated with the $j$-th source. Since all the sources are independent, the joint likelihood is the product of likelihoods for each individual source. Each likelihood is computed using the p.d.f.s. simulated following the procedure of Section \ref{sec3}. For the null hypothesis, with only EBL, the parameter $\theta$ is set to $\theta_{0}=(m_{a},g_{a\gamma})=(0,0)$. For the alternative hypothesis $\theta_{1}=(m_{a},g_{a\gamma})$ takes values in the ALPs parameter space for which the critical energy remains within the {\it Fermi}-LAT energy range. The joint likelihood function is given by,
\begin{equation}
L(E_{1},E_{2},...,E_{N}|m,g)=\prod_{j=1}^{N}f_{j}(E_{j},m,g).
\end{equation}
With the joint likelihood functions we can define different test-statistics (TS) and perform a statistical hypothesis testing between models. We use the following TS, $\Lambda$, defined as the log-likelihood ratio test
\begin{equation}
\Lambda(E_{1},E_{2}...E_{N})=2\log\left(\frac{L(E_{1},E_{2},...,E_{N}|\max\theta_{1})}{L(E_{1},E_{2},...,E_{N}|\theta_{0})}\right),\label{eq34}
\end{equation}
where $\theta_{1}$ now stands for a composite ALPs domain consisting of a set of points within the main region, taken between $0.1\leq m_{a}\leq10$ neV and $0.5\leq g_{11}\leq7.0$. In order to draw any conclusions, we need to compare the observed value of TS, $\Lambda_{obs}$, to different acceptance or rejection thresholds integrated from the null and alternative TS distributions. The null $\Lambda$ distribution, $f_{\Lambda}(H_{0})$, is derived by generating $10^{4}$ Monte-Carlo events under the HEP probability distributions simulated with the null hypothesis. We compute the detection threshold $\Lambda_{thr}$ by integrating 95\% of this distribution. If $\Lambda_{obs}<\Lambda_{thr}$, observations are compatible with the null hypothesis. On the contrary, if $\Lambda_{obs}>\Lambda_{thr}$, the TS is too large to be compatible with the null hypothesis and we could claim a $2\sigma$ significance signal discovery. The alternative TS distributions, $f_{\Lambda}(H_{1}(m_{a},g_{a\gamma})),$ are derived in the same way, but the events are now simulated under the alternative hypotheses. The exclusion thresholds for the ALPs hypotheses, $\Lambda_{exc}(m_{a},g_{a\gamma})$, are computed by integrating 5\% of these distributions. If $\Lambda_{obs}<\Lambda_{exc}(m_{a},g_{a\gamma})$, the set of ALPs parameters would be rejected. This sub-domain of parameters is chosen\emph{ a posteriori}, testing different sets of points until the medians of the alternative distributions are larger than the rejection thresholds for the null hypothesis. 

\section{\label{sec5}Results and conclusions}

\begin{figure}[H]
\centering 
\includegraphics[width=0.8\textwidth]{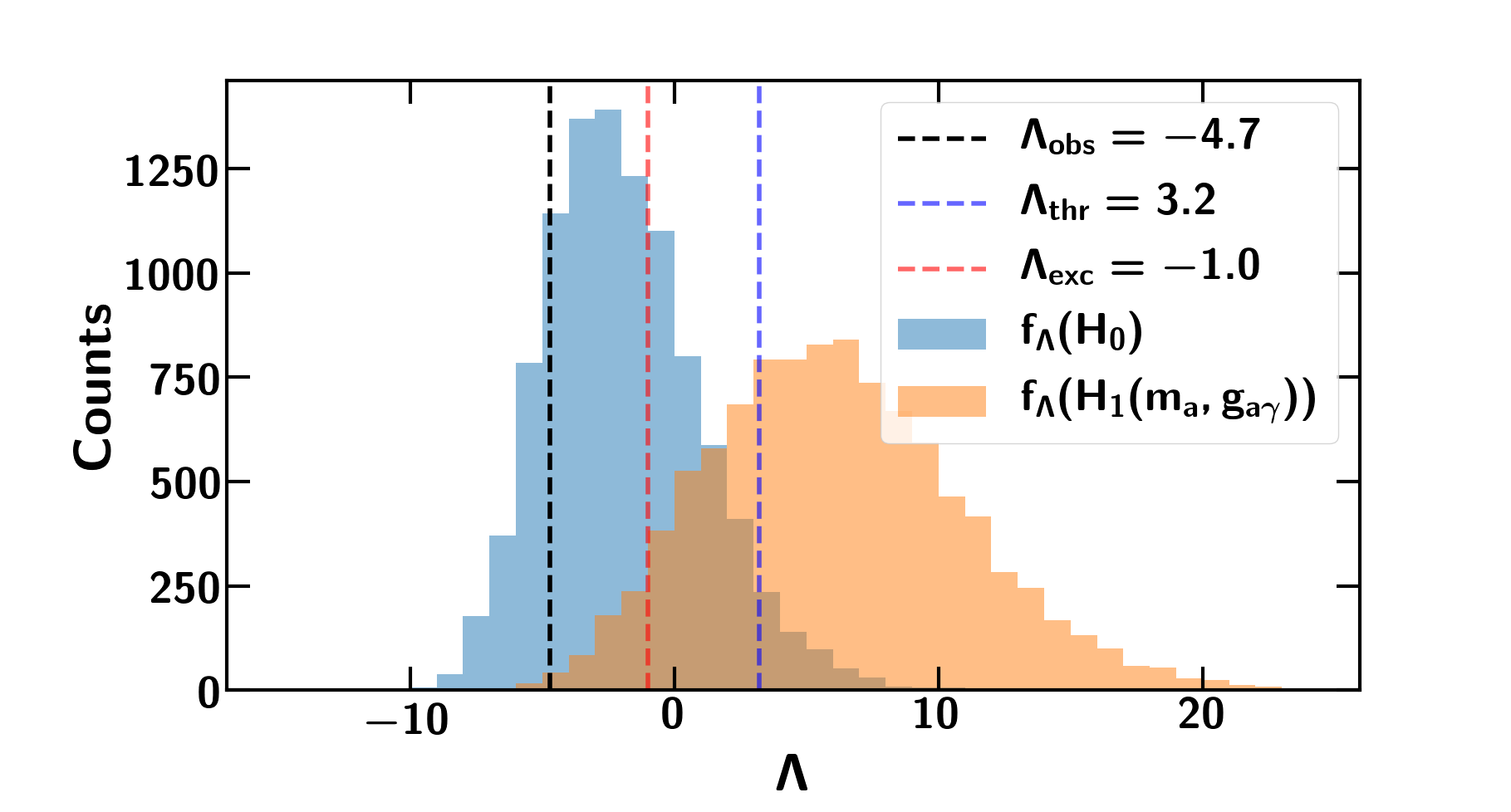} 
\caption[Results]{The simulated null TS distribution (blue) compared to the simulated  alternative TS distribution (orange) for $m_a = 1.3$ neV and $g_{11} =5.2$. Dashed black line: observed value of TS. Dashed blue line: $2\sigma$ detection threshold. Dashed red line: 95$\%$ confidence exclusion threshold.}     
\label{fig:dist}  
\end{figure}

The null and alternative TS distributions, $f_{\Lambda}(H_{0})$ and $f_{\Lambda}(H_{1}(m_{a},g_{a\gamma}))$ for $m_{a}=1.3$ neV and $g_{11}=5.2$ respectively, are shown in Fig. \ref{fig:dist}. The observed TS found with the HEP data is shown as well, $\Lambda_{obs}=-4.7$. The $2\sigma$ detection threshold, derived by integrating the null distribution, is $\Lambda_{thr}=3.2$. As can be seen, $\Lambda_{obs}<\Lambda_{thr}$, therefore the results are compatible with the null hypothesis and no evidence for ALPs was found in these data. An upper limit is set by computing the 95\% exclusion thresholds $\Lambda_{exc}(m_{a},g_{a\gamma})$, also displayed in Fig. \ref{fig:dist}, and testing whether $\Lambda_{obs}<\Lambda_{exc}(m_{a},g_{a\gamma})$. This is repeated for each point tested in the sub-domain.

\begin{figure}[H]
\centering 
\includegraphics[width=0.9\textwidth]{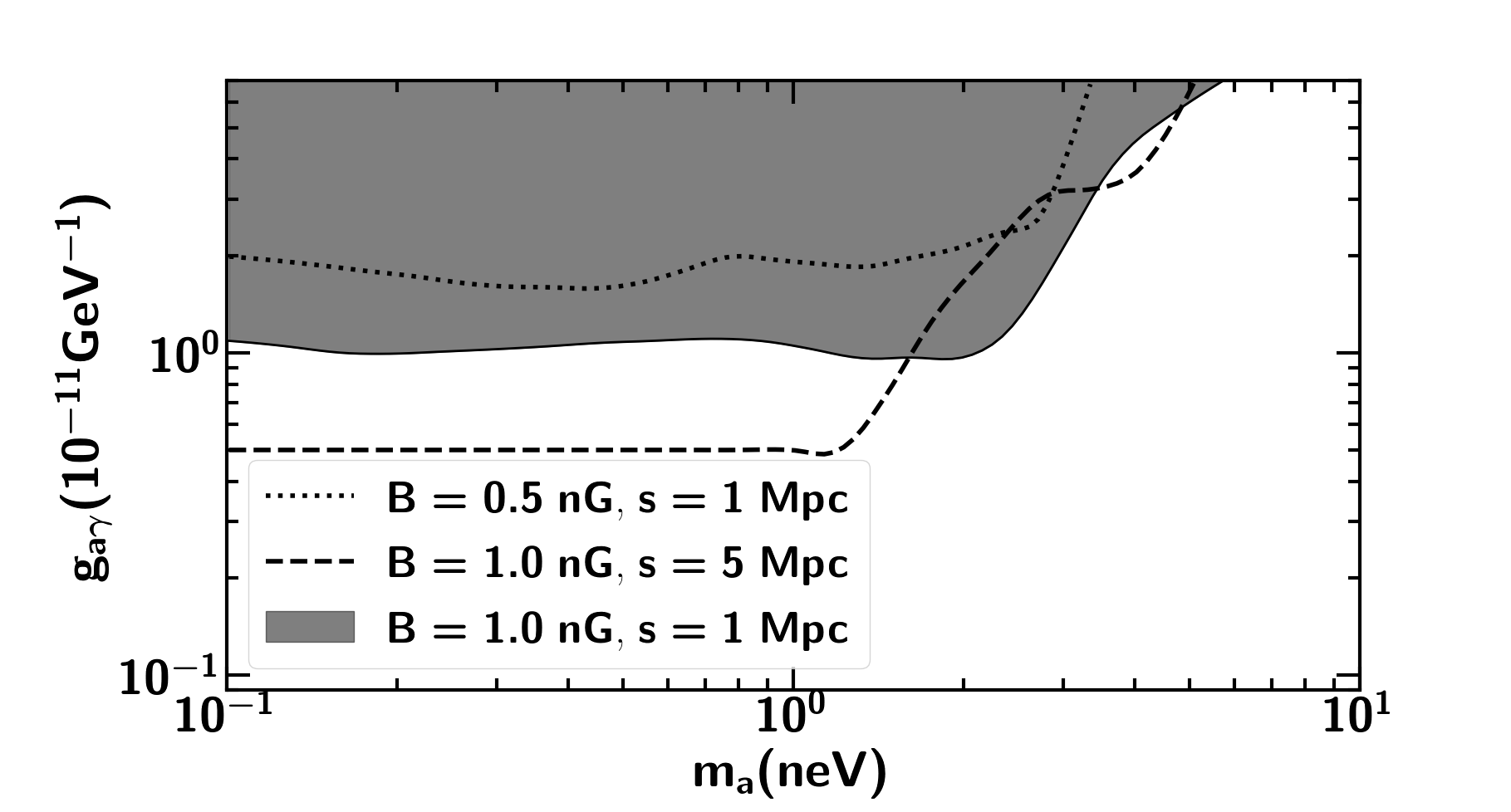} 
\caption[Results]{Shaded region: constraints on the ALP mass-coupling parameter space derived with $B=1$ nG and $s=1$ Mpc. Dotted line: results derived with $B=0.5$ nG and $s=1$ Mpc. Dashed line: results derived with $B=1$ nG and $s=5$ Mpc. This analysis was not sensitive to $B=0.1$ nG and $s=1$ Mpc below $g_{11}$ = 7.0.}     
\label{fig:sys:bf}  
\end{figure}

\begin{figure}[tbp]
\centering 
\includegraphics[width=1\textwidth]{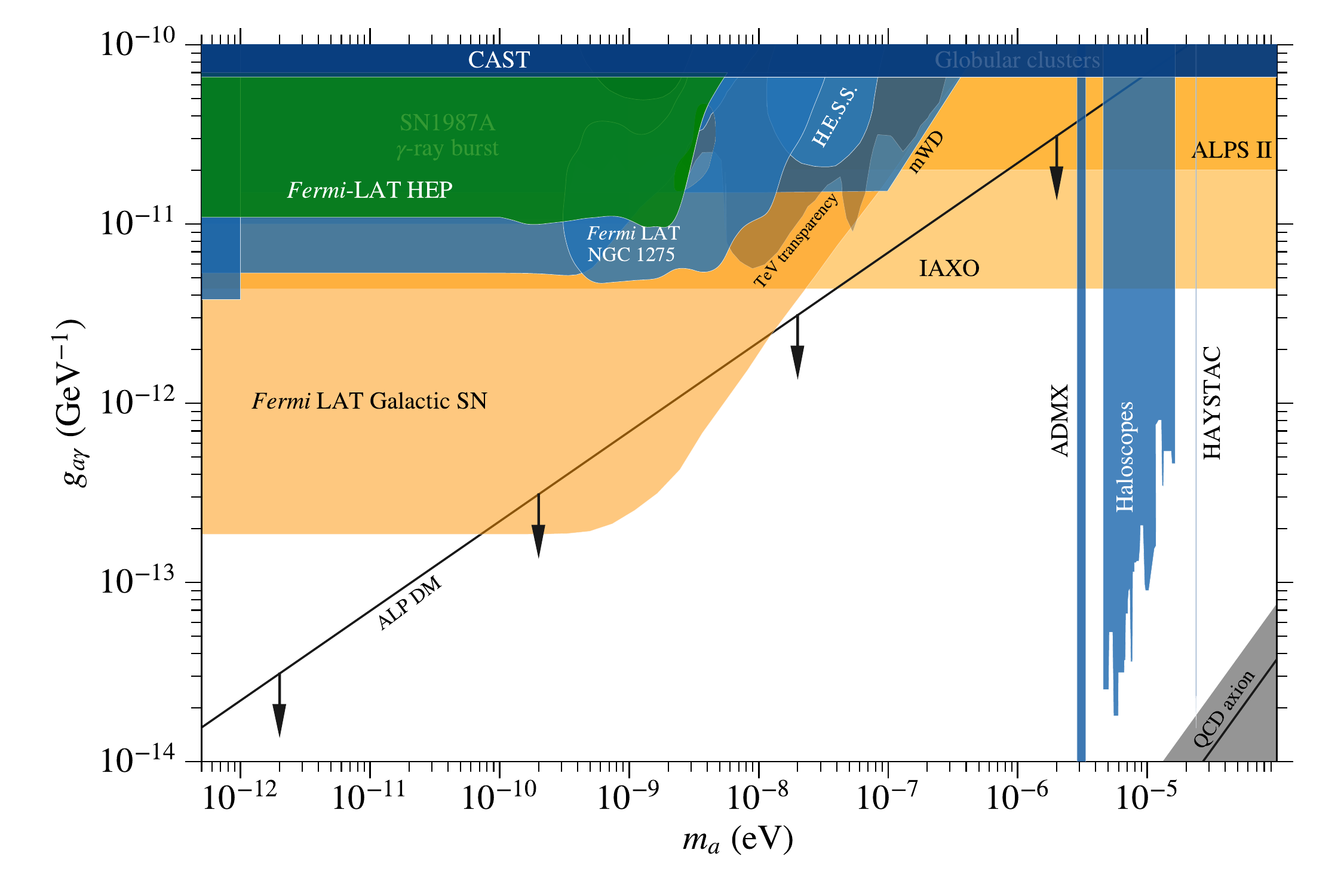}
\hfill
\caption{\label{fig:results} Green: 95$\%$ confidence exclusion region derived in this work with $B=1$ nG and $s=1$ Mpc. Blue: limits from other experiments. Orange: sensitivities for future experiments. Grey: QCD axion. Below the black dashed line ALPs are candidates for the totality of cold dark matter in the Universe.       }
\end{figure}

The resulting upper limits can be seen in Figs. \ref{fig:sys:bf} and \ref{fig:results}. For $B_{T}=1$ nG and $s=1$ Mpc, photon-ALP couplings between $1.0$$\lesssim g_{11}\lesssim7.0$ are excluded for masses below $m_{a}$ $\lesssim3.0$ neV. On the right side,  the contour follows the constant critical energy diagonal from Eq. \ref{eq4}, whereas the horizontal line around $g_{11}\sim1$ depends upon the product $B_{T}\cdot s$. This horizontal line extends to arbitrarily small masses since $B_{T}\cdot s$ does not depend on $m_{a}$. When $m_{a}^{2}<\omega_{pl,neV}^{2}$, the effective mass takes the value of the plasma frequency of the medium. The fluctuations in the contours are due to the limited number of magnetic field realizations and pseudo experiments in the simulation.

The dominant systematic uncertainties are related to the choice of magnetic field parameters. Only upper limits exist for the strength and coherence length of the IGMF \cite{IGM:ns}. We repeat the simulation and likelihood analysis by decreasing the field strength to $B=0.5$ nG and $B=0.1$ nG. The former value yields upper limits that are smaller in  excluded region area by $\sim30\%$ compared to the initial case, while for the latter we cannot set any upper limits below $g_{11}$ = 7.0, as the effects on the transparency are very small. This value is slightly above the CAST limit and couplings below can only be ruled out by magnetic field parameters such that $Bs\gtrsim0.1$. For lower values of B and s,  couplings above the CAST limit can also be excluded. We also increase the coherence length to $s=5$ Mpc. In this scenario, the resulting upper limits increase by roughly $\sim30\%$. These results, seen in Fig. \ref{fig:sys:bf}, are consistent with the ALPs mixing equations. Other sources of systematic uncertainties, such as the EBL model, energy dispersion effects, a different set of magnetic field realizations and HEP probability cuts are smaller than $\sim15\%$ and are discussed in Appendix B.

The limits derived in this work are compatible with other limits and sensitivities of future ALPs experiments, as shown in Fig. \ref{fig:results}. Together with the SN 1987A $\gamma$ burst experiment \cite{AP:supernovae} and with previous {\it Fermi}-LAT \cite{TheFermi-LAT:2016zue} results, our limits strongly constrain part of the parameter space in which ALPs can modify the opacity of the Universe to $\gamma$ rays. Our limits also constrain a part of the unexplored parameter space that the {\it Fermi}-LAT work using observations of the central AGN in the Perseus galaxy cluster could not cover before, i.e., the hole around $g_{11}\sim3$ and $m_{a}\sim3$ neV. They are also within the planned sensitivities of ALPS II \cite{alps2s} and IAXO \cite{iaxos}. None of these limits constrain the region where ALPs could compose the entirety of dark matter content of the Universe \cite{alps:cdm_mm} which is below the dashed black line in Fig. \ref{fig:results}.

Magnetic field morphologies in the interstellar and intergalactic space are not fully understood yet. Better observations of cosmic magnetic fields are needed in order to reduce the systematic uncertanties associated to these fields, which are crucial for the photon-ALP beam propagation. Future experiments like JVLA \cite{jvla}, ALMA \cite{alma}, and SKA \cite{ska} will be able to improve these current limitations \cite{FIELDS:rev}. Recent EBL results, such as Ref. \cite{desai19}, may also be used in future ALPs analyses.

A part of the parameter space where ALPs could affect the transparency of the Universe to $\gamma$ rays remains to be explored. For larger values of $m_{a}$, the critical energy increases and the maximal conversion probability takes place in the TeV range. The conversion is further enhanced by the inclusion of the galactic magnetic field in the propagation of the beam. Current Cherenkov telescopes, with energy ranges within the TeV range can reach higher masses in the parameter space and improve the limits derived in this work. The future Cherenkov Telescope Array (CTA) \cite{cta_2016} can probe even higher masses. Hence, the combined likelihood analysis of many sources presented in this work can be extended to these instruments. The number of events within a given energy range can be measured and compared to simulations that include ALPs models. Other magnetic field scenarios may also be included to improve the results. Recently, indications for ALPs have been presented in Refs. \cite{ind:ciber,ind:gamma}.  In contrast to our HEP analysis, they use different samples of sources at different redshifts and higher energies,  different observables and other EBL models. In the former, authors use data from the CIBER collaboration that indicates a higher EBL density, thus making current $\gamma$-ray observations compatible with ALPs over a certain region of the parameter space. Part of this region has already been excluded by other experiments and recent EBL results are not in tension with previous models \cite{desai19}. Authors from Ref. \cite{ind:gamma} found evidence for an evolution of the spectral indices with redshift of blazars observed with MAGIC, HESS and VERITAS\footnote{\url{https://www.mpi-hd.mpg.de/hfm/HESS/}

\url{https://magic.mpp.mpg.de/}

\url{https://veritas.sao.arizona.edu/}}. They argue that such an effect is not expected in a purely astrophysical scenario and indeed find that the effect disappears when photon-ALP oscillations are included in their analysis. They find a best-fit parameter $\xi = (B/nG) (g_{a\gamma}/10^{-11}GeV^{-1}) = 0.5$. With our current analysis, we can rule out $\xi\geq 1$. All these predictions can be improved and tested with more accurate magnetic field and EBL models and even current Cherenkov telescopes and the upcoming CTA.

\acknowledgments

The \textit{Fermi} LAT Collaboration acknowledges generous ongoing support from a number of agencies and institutes that have supported both the development and the operation of the LAT as well as scientific data analysis. These include the National Aeronautics and Space Administration and the Department of Energy in the United States, the Commissariat \`a l'Energie Atomique and the Centre National de la Recherche Scientifique / Institut National de Physique Nucl\'eaire et de Physique des Particules in France, the Agenzia Spaziale Italiana and the Istituto Nazionale di Fisica Nucleare in Italy, the Ministry of Education, Culture, Sports, Science and Technology (MEXT), High Energy Accelerator Research Organization (KEK) and Japan Aerospace Exploration Agency (JAXA) in Japan, and the K.~A.~Wallenberg Foundation, the Swedish Research Council and the Swedish National Space Board in Sweden.   Additional support for science analysis during the operations phase is gratefully acknowledged from the Istituto Nazionale di Astrofisica in Italy and the Centre National d'\'Etudes Spatiales in France. This work performed in part under DOE Contract DE-AC02-76SF00515.

This work was possible thanks to the DESY Strategy Fund program.

Alberto Dom\'inguez thanks the support of the  Ram\'on y Cajal program from the Spanish MINECO.

\appendix
\section{\label{AA} Survival probability}

In this Appendix we provide a brief summary of the survival probability derivation, which is treated in detail in Ref. \cite{GOLD:formulas}. From Eq. \ref{eq1}, the photon-ALP propagation for a monochromatic system along the $y$-axis, in case that the photon energy is $E\gg m_{a}$, can be described by \cite{oscillations:raffelt}

\begin{equation}
\left(i\partial_{y}+E+\mathcal{M}_{0}\right)\left(\begin{array}{c}
A_{x}(y)\\
A_{z}(y)\\
a(y)
\end{array}\right)=0,\label{eq:beam}
\end{equation}
where $E$ is the photon energy, $\mathcal{M}_{0}$ is the mixing matrix and the second factor is $\psi(y)$, the beam state vector. The photon polarization amplitudes along the $x$- and $z$-axis are denoted by $A_{x}(y)$ and $A_{z}(y)$, respectively, while the ALP field amplitude is $a(y)$. The mixing matrix is real and symmetric, and it involves different terms. The general form is given by:
\begin{equation}
\mathcal{M}_{0}=\left(\begin{array}{ccc}
\Delta_{xx} & \Delta_{xz} & \Delta_{a\gamma}^{x}\\
\Delta_{zx} & \Delta_{zz} & \Delta_{a\gamma}^{z}\\
\Delta_{a\gamma}^{x} & \Delta_{a\gamma}^{z} & \Delta_{aa}
\end{array}\right).
\end{equation}

The $\Delta_{a\gamma}$-terms and the $\Delta_{aa}$-term represent the mixing of photons with ALPs from Eq. \ref{eq1} and ALPs self-interactions, respectively. The remaining terms depend on the properties of the medium, namely QED vacuum effects and absorption mechanisms. The former come from the Heisenberg-Euler-Weisskopf (HEW) effective Lagrangian for the photon one-loop vacuum polarization under an external magnetic field \cite{HSW:lag}. Vacuum QED terms can be ignored at high energies. The other contribution is due to background particles in the medium that may annihilate the primary photon, such as EBL.

The photon-ALP beam has to be treated as unpolarized, therefore it is described by a polarization density matrix, 
\begin{equation}
\rho(y)=\psi(y)\otimes\psi(y)^{\dagger},
\end{equation}
which obeys the Von Neumann equation of non-relativistic quantum mechanics
\cite{GOLD:formulas}, 
\begin{equation}
i\partial_{y}\rho=\left[\rho,\mathcal{M}\right]=\rho\mathcal{M}^{\dagger}-\mathcal{M}\rho,
\end{equation}

The solution of the propagation of the beam is given by the transfer
matrix, 
\begin{equation}
\rho(y)=\mathcal{U}(y,y_{0})\rho(y_{0})\mathcal{U^{\dagger}}(y,y_{0}).\label{eq:5.5}
\end{equation}

The transition probability from one state to another state is given by the trace of the projection of both states, 
\begin{equation}
P_{\rho_{1}\rightarrow\rho_{2}}=\mathrm{Tr}\left(\rho_{2}\mathcal{U}(y,y_{0})\rho_{1}\mathcal{U^{\dagger}}(y,y_{0})\right).\label{eq:prob_transfer}
\end{equation}

For the intergalactic medium, the magnetic region is split into $N$ domains with a homogeneous field in each one. For a homogeneous field, the problem is simplified because we can always align $\mathbf{B_{T}}$ with the $z$-axis. In this case, this can no longer be done across all the domains, but the problem can still be simplified through similarity transformations, 
\begin{equation}
\mathcal{M}=V^{\dagger}(\phi)\mathcal{M}_{0}V(\phi),
\end{equation}
where $V(\phi)$ is the rotation matrix in the $x-z$ plane perpendicular to the propagation direction and $\phi$ is the angle $\mathbf{B_{T}}$ forms with the $z$-axis in each domain. The full transfer matrix of the system across $N$ domains is, 
\begin{equation}
\mathcal{U}(E_{0},z,\phi_{1}...\phi_{N})=\prod_{i=1}^{N}\mathcal{U}(E_{i},\phi_{i})\label{eq:transfer_matrix}
\end{equation}

The transition probability between two states is computed with Eqs. \ref{eq:prob_transfer} and \ref{eq:transfer_matrix}. The propagation is a stochastic process due to the random orientations of the field in each domain, therefore only the mean properties of the beam can be evaluated. Moreover, we have to sum over the two final polarization states because of the lack of polarization measurements. The photon
survival probability is,
\begin{equation}
P_{\gamma\rightarrow\gamma}(E_{0},z)=\sum_{i=1,2}\mathrm{Tr}\left(\left\langle \rho_{i}\mathcal{U}(E_{0},z,\phi_{1}...\phi_{N})\rho_{\mathrm{unpol}}\mathcal{U}^{\dagger}(E_{0},z,\phi_{1}...\phi_{N})\right\rangle \right)_{\phi_{1}...\phi_{N}}\label{eq:general_pag}
\end{equation}

The modified photon survival probabilities for each grid point are computed with this equation.

\section{\label{AB} Other uncertainties}

In order to evaluate uncertainties associated to model parameters and analysis choices, we derive different sets of limits repeating the simulation and analysis procedures of Sections \ref{sec3} and \ref{sec4}. For each test, we extract the percentage changed in area from one case to another for the particular grid explored in this study, between $0.1\leq m_{a}\leq10$ neV and $0.5\leq g_{11}\leq7.0$. We refer as area the visual region in logarithmic scale that is excluded for each case. 

\begin{figure}[H]
\centering 
\includegraphics[width=0.6\textwidth]{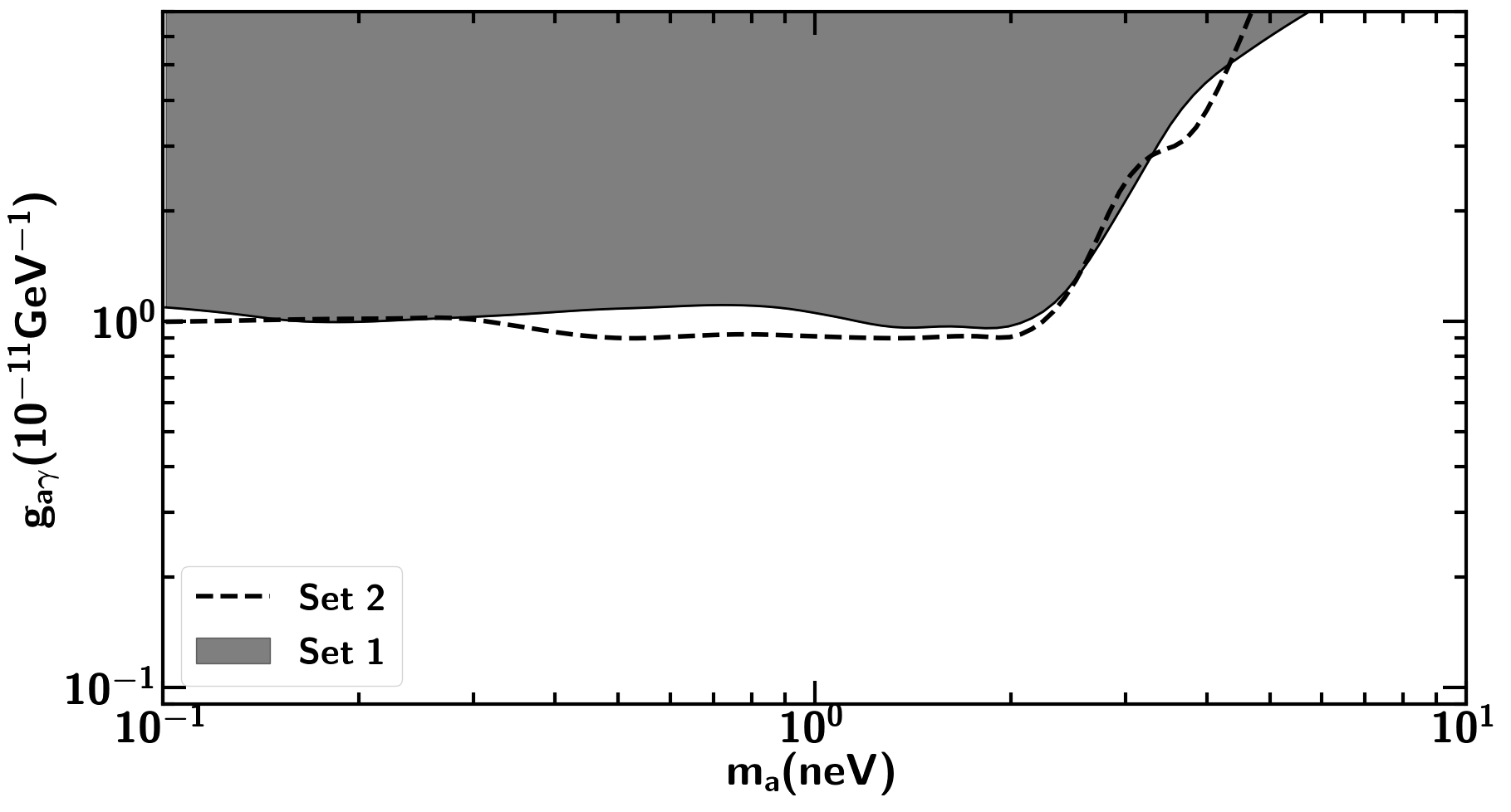}
\caption[Results]{Shaded region: constraints on the ALP mass-coupling parameter space derived with the average of a set of magnetic field realizations. Dashed line: results derived with a different set of realizations.  }     
\label{fig:sys:vi}  
\end{figure}

\begin{figure}[H]
\centering 
\includegraphics[width=0.6\textwidth]{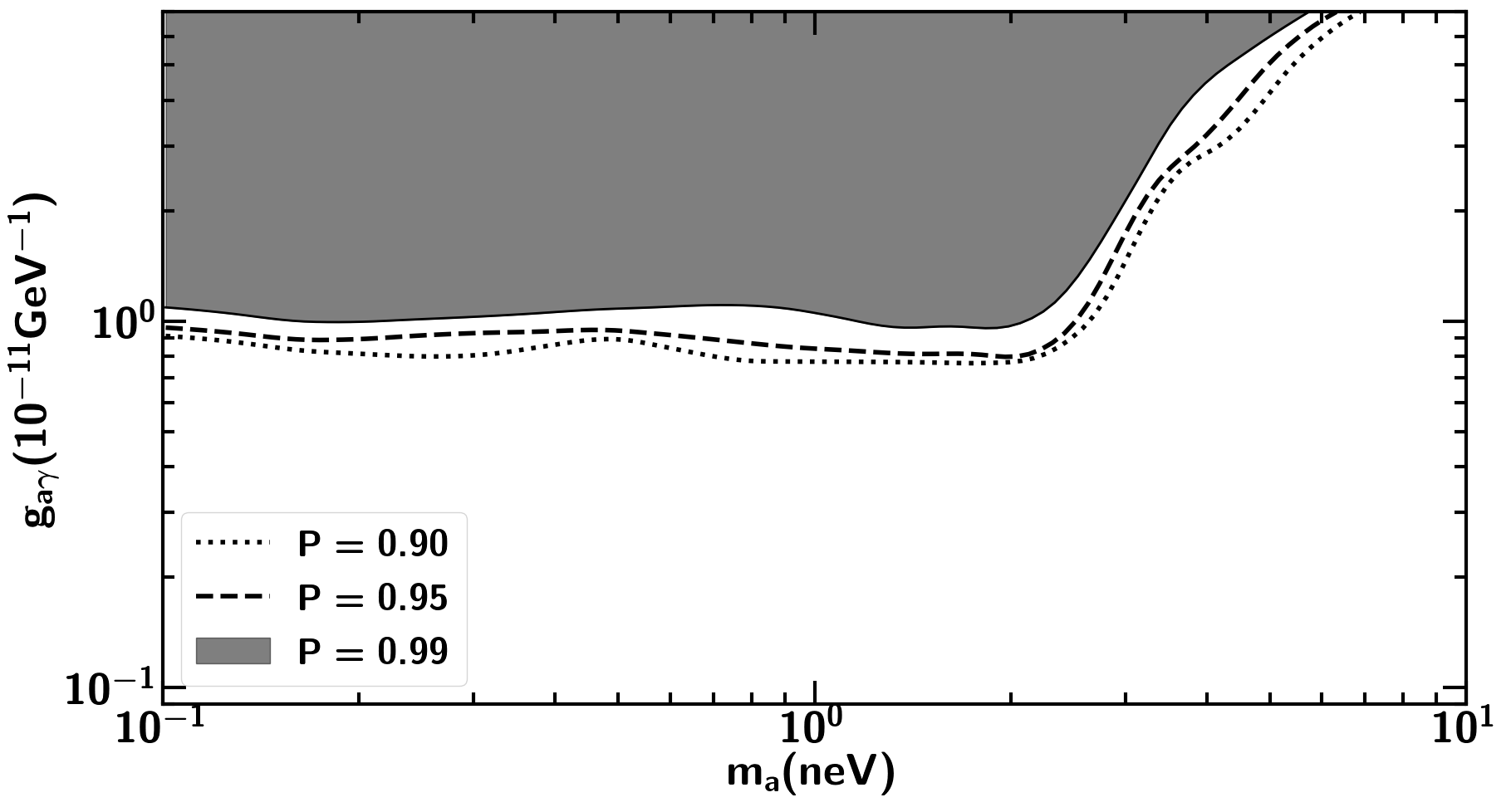}
\caption[Results]{Shaded region: constraints on the ALP mass-coupling parameter space derived with $P=0.99$. Dashed line: results derived with $P=0.95$. Dotted line: Results derived with $P=0.90$.  }     
\label{fig:sys:pi}  
\end{figure}

Equation \ref{eq:general_pag} computes the average survival probability over IGMF realizations along the line of sight of each source. The oscillating contours from Fig. \ref{fig:results} come from a limited number of simulations and field realizations. These two effects are tested for a different set of pseudo experiments and field realizations, resulting in the limits of Fig. \ref{fig:sys:vi}. The exclusion region changes are smaller than $\sim10\%$.
In Section \ref{sec3}, we discussed the AGN data sample and took sources based on the HEP probability to belong to a source. All of the sources in the 2FHL catalog have a HEP with $P\geq0.85$, with most of them above $P\geq0.99$, due to the low background of the LAT at high energies. Selecting sources with higher values of $P$ allows us to reduce the events that come from background. However, this also entails a reduction of statistics in our sample. We tested the effects of different HEP probability cuts within one realization, resulting in contours with area changes smaller than $\sim10\%$, as displayed in Fig. \ref{fig:sys:pi}.

\begin{figure}[H]
\centering 
\includegraphics[width=0.6\textwidth]{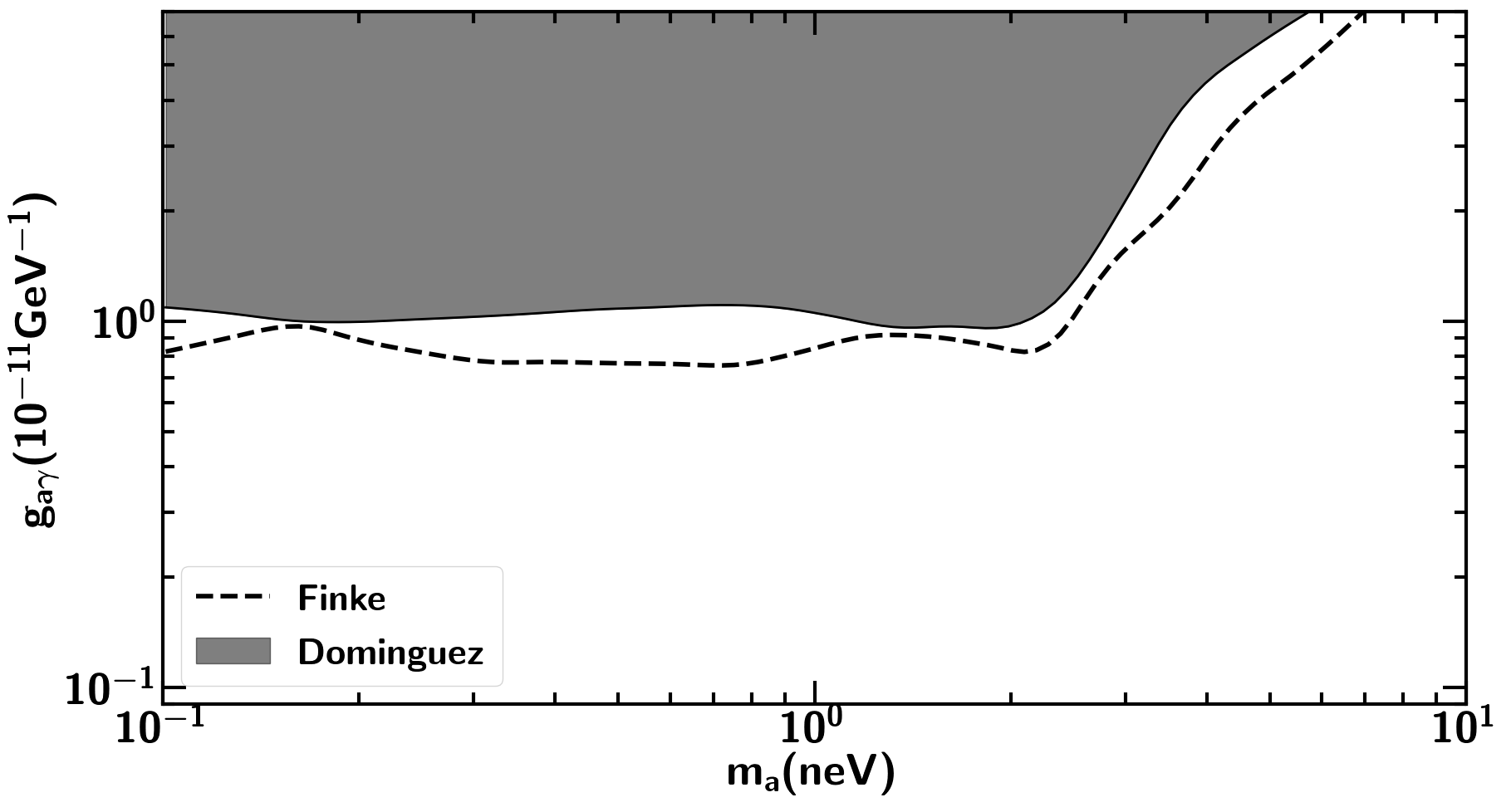}   
\caption[Results]{Shaded region: constraints on the ALP mass-coupling parameter space derived with the Dom\'inguez et al. EBL model. Dashed line: results derived with the Finke et al. EBL model.}     
\label{fig:sys:ebl}  
\end{figure}

Our results were derived with the Dom\'inguez et al. EBL model. We test the effects of choosing a different model by repeating the analysis with the Finke et al. model \cite{EBL:forward1}. The upper limits increase by $\sim15\%$, as seen in Fig. \ref{fig:sys:ebl}. Finally, we did not consider energy dispersion effects. The reason for this is that, above 1 GeV, these effects are below 10\% at 68\% confidence and thus we do not expect a large change in the results.

All the results were computed with the mean values of the spectral fit parameters, which have their own statistical uncertainties that we did not take into account. A way to do this would be to bias the fluxes of the sources to produce more photons at high energies, then the probability distributions would shift to the right, changing the likelihood values for both models and all sources. We did not perform this for two reasons. First, this is similar to modifying the opacity of the Universe with a different propagation model, which we already carried out with the Finke et al. model, resulting in a limits difference that is still smaller than the dominant uncertainty imposed by the magnetic field. Second, the parameters of the spectra would take random values for each source that might cancel out in the combined likelihood analysis of all the sources. Additionally, some of these spectra accept different model representations and are not bound to only power laws or only logarithmic parabolas, hence we tested the effects of choosing different sets of parameters for the sources. The results were still smaller than 10\% compared to the main limits we obtained in this work. Finally, the effective areas have also associated systematic uncertainties at high energies. It is usually evaluated through a bracketing procedure during the LAT data analysis. Its effects at the energies considered in our analysis are estimated to yield differences in photon counts between 5\% and 15\%. This is still below the dominant systematic uncertainty from the magnetic field. 

\section{\label{AC}Sources and fits}

Table \ref{tab:srcs} summarizes the spectral fits for all the sources in the 2FHL catalog with $z\geq0.1$. The first column is the spectral shape, which can be a power law (pow) or a logarithmic parabola (log). The second column is the $\chi^{2}$ value from the minimization procedure. The third column, ndf, is the number of degrees of freedom. The last column denotes the p-value.

Table \ref{tab:sfits4} summarizes the fit parameters for all the sources with a pivot energy of $100$ GeV. The column indices follow the logarithmic parabola equation parameters and its corresponding errors:
\begin{equation}
\frac{dN}{dE}=K\left(\frac{E}{E_{0}}\right)^{-\left(G+A\log\left(E/E_{0}\right)\right)}
\end{equation}


\begin{table}[H]
  \small
  \centering
  \begin{tabular}{|c|c|c|c|c|}
    \hline
    \textbf{Source} & \textbf{Shape} & \textbf{$\chi^2$} & \textbf{ndf} & \textbf{p-val}\\
    \hline 
    \input{Tables/fit_q1.tex}
    \hline
  \end{tabular}
\end{table}

\begin{table}[H]
  \small
  \centering
  \begin{tabular}{|c|c|c|c|c|}
    \hline
    \textbf{Source} & \textbf{Shape} & \textbf{$\chi^2$} & \textbf{ndf} & \textbf{p-val}\\
    \hline 
    \input{Tables/fit_q2.tex}
    \hline
  \end{tabular} 
\end{table}

\begin{table}[H]
  \small
  \centering
  \begin{tabular}{|c|c|c|c|c|}
    \hline
    \textbf{Source} & \textbf{Shape} & \textbf{$\chi^2$} & \textbf{ndf} & \textbf{p-val}\\
    \hline 
    \input{Tables/fit_q3.tex}
    \hline
  \end{tabular} 
\end{table}

\begin{table}[H]
  \small
  \centering
  \begin{tabular}{|c|c|c|c|c|}
    \hline
    \textbf{Source} & \textbf{Shape} & \textbf{$\chi^2$} & \textbf{ndf} & \textbf{p-val}\\
    \hline 
    \input{Tables/fit_q4.tex}
    \hline
  \end{tabular}
  \caption[DM Halo TS Values]{       
	Fit results for the 2FHL catalog sources with redshifts $z\geq 0.1$.    
  }
  \label{tab:srcs}   
\end{table}

\begin{table}[H]
  \small
  \centering
  \begin{tabular}{|c|c|c|c|c|c|c|}
    \hline
    \textbf{Source} & \textbf{$G$} & \textbf{$errG$} & \textbf{$A$} & \textbf{$errA$}& \textbf{$K$} & \textbf{$errK$}\\
    \hline 
    \input{Tables/fit_p1.tex}
    \hline
  \end{tabular} 
\end{table}

\begin{table}[H]
  \small
  \centering
  \begin{tabular}{|c|c|c|c|c|c|c|}
    \hline
    \textbf{Source} & \textbf{$G$} & \textbf{$errG$} & \textbf{$A$} & \textbf{$errA$}& \textbf{$K$} & \textbf{$errK$}\\
    \hline 
    \input{Tables/fit_p2.tex}
    \hline
  \end{tabular} 
\end{table}

\begin{table}[H]
  \small
  \centering
  \begin{tabular}{|c|c|c|c|c|c|c|}
    \hline
    \textbf{Source} & \textbf{$G$} & \textbf{$errG$} & \textbf{$A$} & \textbf{$errA$}& \textbf{$K$} & \textbf{$errK$}\\
    \hline 
    \input{Tables/fit_p3.tex}
    \hline
  \end{tabular}  
\end{table}

\begin{table}[H]
  \small
  \centering
  \begin{tabular}{|c|c|c|c|c|c|c|}
    \hline
    \textbf{Source} & \textbf{$G$} & \textbf{$errG$} & \textbf{$A$} & \textbf{$errA$}& \textbf{$K$} & \textbf{$errK$}\\
    \hline 
    \input{Tables/fit_p4.tex}
    \hline
  \end{tabular}
  \caption[DM Halo TS Values]{       
	Fit parameters for the 2FHL catalog sources with redshifts $z\geq 0.1$.    
  }
  \label{tab:sfits4}   
\end{table}


\bibliography{thesis_bib} 
\bibliographystyle{ieeetr}

\end{document}

%% file: Tables/fit_q1.tex
2FHLJ0008.1$+$4709 & pow & 3.213 & 2.0 & 0.201 \\
2FHLJ0022.0$+$0006 & pow & 0.985 & 2.0 & 0.611 \\
2FHLJ0033.6$-$1921 & log & 1.461 & 2.0 & 0.482 \\
2FHLJ0043.9$+$3424 & log & 0.956 & 2.0 & 0.62 \\
2FHLJ0114.9$-$3359 & log & 0.278 & 2.0 & 0.87 \\
2FHLJ0115.8$+$2519 & pow & 0.648 & 3.0 & 0.885 \\
2FHLJ0123.7$-$2308 & pow & 0.166 & 2.0 & 0.921 \\
2FHLJ0141.3$-$0927 & pow & 0.983 & 3.0 & 0.805 \\
2FHLJ0221.1$+$3556 & log & 1.48 & 2.0 & 0.477 \\
2FHLJ0222.6$+$4301 & log & 1.43 & 2.0 & 0.489 \\
2FHLJ0237.6$-$3605 & pow & 1.472 & 2.0 & 0.479 \\
2FHLJ0238.4$-$3116 & pow & 6.716 & 4.0 & 0.152 \\
2FHLJ0238.8$+$1631 & log & 1.454 & 1.0 & 0.228 \\
2FHLJ0244.9$-$5820 & pow & 10.351 & 4.0 & 0.035 \\
2FHLJ0303.3$-$2407 & pow & 7.601 & 3.0 & 0.055 \\
2FHLJ0304.5$-$0054 & pow & 0.453 & 2.0 & 0.797 \\
2FHLJ0316.1$+$0905 & log & 1.042 & 1.0 & 0.307 \\
2FHLJ0319.7$+$1849 & pow & 1.018 & 3.0 & 0.797 \\
2FHLJ0326.0$-$1644 & pow & 3.981 & 3.0 & 0.264 \\
2FHLJ0326.3$+$0227 & pow & 2.419 & 3.0 & 0.49 \\
2FHLJ0416.9$+$0105 & pow & 0.778 & 3.0 & 0.855 \\
2FHLJ0422.9$+$1947 & pow & 2.702 & 2.0 & 0.259 \\
2FHLJ0428.7$-$3755 & log & 2.528 & 1.0 & 0.112 \\
2FHLJ0433.6$+$2907 & log & 1.692 & 2.0 & 0.429 \\

%% file: Tables/fit_q2.tex
2FHLJ0440.3$-$2458 & log & 0.024 & 1.0 & 0.877 \\
2FHLJ0449.4$-$4349 & log & 1.831 & 1.0 & 0.176 \\
2FHLJ0456.9$-$2323 & log & 0.463 & 1.0 & 0.496 \\
2FHLJ0507.9+6737 & pow & 4.778 & 2.0 & 0.092 \\
2FHLJ0538.6$-$4406 & log & 0.093 & 1.0 & 0.76 \\
2FHLJ0543.9$-$5533 & log & 0.82 & 2.0 & 0.664 \\
2FHLJ0622.4$-$2604 & pow & 5.604 & 2.0 & 0.061 \\
2FHLJ0631.0$-$2406 & log & 1.406 & 2.0 & 0.495 \\
2FHLJ0648.6+1516 & log & 0.26 & 3.0 & 0.967 \\
2FHLJ0650.7+2502 & log & 0.498 & 2.0 & 0.78 \\
2FHLJ0710.5+5908 & log & 2.375 & 3.0 & 0.498 \\
2FHLJ0721.9+7121 & log & 0.144 & 2.0 & 0.93 \\
2FHLJ0744.2+7435 & pow & 1.828 & 3.0 & 0.609 \\
2FHLJ0756.8+0955 & pow & 3.504 & 3.0 & 0.32 \\
2FHLJ0805.8+7534 & log & 1.084 & 3.0 & 0.781 \\
2FHLJ0809.7+5218 & log & 1.301 & 3.0 & 0.729 \\
2FHLJ0811.6+0146 & log & 2.548 & 2.0 & 0.28 \\
2FHLJ0825.9$-$2230 & pow & 2.181 & 2.0 & 0.336 \\
2FHLJ0847.2+1133 & pow & 5.598 & 3.0 & 0.133 \\
2FHLJ0950.2+4553 & pow & 3.59 & 2.0 & 0.166 \\
2FHLJ0952.2+7503 & log & 0.27 & 1.0 & 0.603 \\
2FHLJ0957.6+5523 & log & 2.064 & 2.0 & 0.356 \\
2FHLJ0958.3+6535 & log & 0.983 & 1.0 & 0.322 \\
2FHLJ1010.2$-$3119 & pow & 1.132 & 2.0 & 0.568 \\

%% file: Tables/fit_q3.tex
2FHLJ1015.0+4926 & log & 3.0 & 2.0 & 0.223 \\
2FHLJ1031.2+7441 & pow & 3.358 & 4.0 & 0.5 \\
2FHLJ1031.4+5052 & log & 3.7 & 3.0 & 0.296 \\
2FHLJ1053.5+4930 & pow & 7.557 & 5.0 & 0.182 \\
2FHLJ1058.5+5625 & log & 5.509 & 4.0 & 0.239 \\
2FHLJ1104.0$-$2331 & pow & 2.126 & 3.0 & 0.547 \\
2FHLJ1116.9+2014 & pow & 5.681 & 4.0 & 0.224 \\
2FHLJ1120.8+4212 & log & 1.971 & 3.0 & 0.579 \\
2FHLJ1125.6$-$3555 & pow & 3.135 & 4.0 & 0.536 \\
2FHLJ1136.8+6737 & log & 1.441 & 4.0 & 0.837 \\
2FHLJ1137.9$-$1710 & log & 0.524 & 1.0 & 0.469 \\
2FHLJ1217.9+3006 & log & 3.382 & 3.0 & 0.336 \\
2FHLJ1221.3+3009 & log & 6.104 & 3.0 & 0.107 \\
2FHLJ1224.4+2435 & log & 2.482 & 3.0 & 0.479 \\
2FHLJ1224.7+2124 & log & 3.409 & 2.0 & 0.182 \\
2FHLJ1231.7+2848 & log & 5.409 & 3.0 & 0.144 \\
2FHLJ1256.2$-$0548 & log & 0.909 & 1.0 & 0.34 \\
2FHLJ1309.5+4305 & log & 0.029 & 1.0 & 0.865 \\
2FHLJ1312.6+4828 & log & 0.462 & 2.0 & 0.794 \\
2FHLJ1404.9+6555 & log & 1.364 & 2.0 & 0.506 \\
2FHLJ1418.0+2543 & pow & 0.782 & 3.0 & 0.854 \\
2FHLJ1427.3$-$4204 & log & 1.192 & 1.0 & 0.275 \\
2FHLJ1428.5+4239 & log & 0.146 & 2.0 & 0.929 \\
2FHLJ1437.0+5639 & log & 1.467 & 3.0 & 0.69 \\

%% file: Tables/fit_q4.tex
2FHLJ1442.9+1159 & pow & 0.651 & 2.0 & 0.722 \\
2FHLJ1506.8+0813 & log & 4.451 & 3.0 & 0.217 \\
2FHLJ1512.7$-$0906 & log & 0.121 & 2.0 & 0.942 \\
2FHLJ1517.7+6526 & log & 0.009 & 1.0 & 0.922 \\
2FHLJ1548.7$-$2249 & pow & 0.563 & 3.0 & 0.905 \\
2FHLJ1748.7+7004 & log & 0.414 & 1.0 & 0.52 \\
2FHLJ1801.2+7828 & log & 0.319 & 1.0 & 0.572 \\
2FHLJ1917.7$-$1921 & log & 0.125 & 2.0 & 0.94 \\
2FHLJ1936.9$-$4721 & pow & 1.958 & 3.0 & 0.581 \\
2FHLJ1958.3$-$3011 & log & 1.526 & 2.0 & 0.466 \\
2FHLJ2000.9$-$1749 & pow & 0.044 & 2.0 & 0.978 \\
2FHLJ2016.5$-$0904 & log & 2.258 & 2.0 & 0.323 \\
2FHLJ2116.1+3339 & log & 0.04 & 1.0 & 0.841 \\
2FHLJ2131.4$-$0914 & log & 1.166 & 2.0 & 0.558 \\
2FHLJ2150.3$-$1411 & log & 0.152 & 2.0 & 0.927 \\
2FHLJ2153.1$-$0041 & pow & 0.175 & 2.0 & 0.916 \\
2FHLJ2158.8$-$3013 & log & 2.695 & 2.0 & 0.26 \\
2FHLJ2249.9+3826 & pow & 3.356 & 3.0 & 0.34 \\
2FHLJ2254.0+1613 & log & 1.116 & 2.0 & 0.572 \\
2FHLJ2314.0+1445 & log & 3.726 & 4.0 & 0.444 \\
2FHLJ2324.7$-$4041 & log & 0.36 & 1.0 & 0.548 \\
2FHLJ2329.2+3754 & log & 1.792 & 2.0 & 0.408 \\
2FHLJ2340.8+8014 & log & 1.622 & 2.0 & 0.444 \\
2FHLJ2343.5+3438 & pow & 0.579 & 3.0 & 0.901 \\

%% file: Tables/fit_p1.tex
2FHLJ0008.1$+$4709 & 2.016 & 0.049 & 0 & 0 & 1.976e-11 & 2.05e-12 \\
2FHLJ0022.0$+$0006 & 1.752 & 0.249 & 0 & 0 & 1.34e-12 & 3.7e-13 \\
2FHLJ0033.6$-$1921 & 1.787 & 0.043 & 0.037 & 0.017 & 4.028e-11 & 1.82e-12 \\
2FHLJ0043.9$+$3424 & 1.947 & 0.041 & 0.007 & 0.014 & 3.03e-11 & 1.22e-12 \\
2FHLJ0114.9$-$3359 & 1.345 & 0.108 & 0.014 & 0.061 & 1.54e-12 & 2.2e-13 \\
2FHLJ0115.8$+$2519 & 1.889 & 0.027 & 0 & 0 & 1.368e-11 & 5.4e-13 \\
2FHLJ0123.7$-$2308 & 1.859 & 0.03 & 0 & 0 & 6.59e-12 & 2.7e-13 \\
2FHLJ0141.3$-$0927 & 2.151 & 0.021 & 0 & 0 & 1.449e-11 & 7.1e-13 \\
2FHLJ0221.1$+$3556 & 2.392 & 0.042 & 0.018 & 0.012 & 4.842e-11 & 1.96e-12 \\
2FHLJ0222.6$+$4301 & 2.008 & 0.021 & 0.022 & 0.007 & 1.5949e-10 & 3.17e-12 \\
2FHLJ0237.6$-$3605 & 1.918 & 0.158 & 0 & 0 & 2.79e-12 & 5.7e-13 \\
2FHLJ0238.4$-$3116 & 1.807 & 0.061 & 0 & 0 & 1.194e-11 & 1.35e-12 \\
2FHLJ0238.8$+$1631 & 2.516 & 0.075 & 0.116 & 0.038 & 7.586e-11 & 3.9e-12 \\
2FHLJ0244.9$-$5820 & 1.773 & 0.118 & 0 & 0 & 6.58e-12 & 1.28e-12 \\
2FHLJ0303.3$-$2407 & 1.906 & 0.028 & 0 & 0 & 6.002e-11 & 3.76e-12 \\
2FHLJ0304.5$-$0054 & 1.951 & 0.128 & 0 & 0 & 1.71e-12 & 2.7e-13 \\
2FHLJ0316.1$+$0905 & 1.974 & 0.09 & 0.1 & 0.06 & 2.088e-11 & 2.02e-12 \\
2FHLJ0319.7$+$1849 & 1.833 & 0.053 & 0 & 0 & 6.97e-12 & 5e-13 \\
2FHLJ0326.0$-$1644 & 1.863 & 0.068 & 0 & 0 & 1.212e-11 & 1.22e-12 \\
2FHLJ0326.3$+$0227 & 1.815 & 0.076 & 0 & 0 & 7.99e-12 & 8.3e-13 \\
2FHLJ0416.9$+$0105 & 1.829 & 0.044 & 0 & 0 & 7.81e-12 & 4.7e-13 \\
2FHLJ0422.9$+$1947 & 1.942 & 0.242 & 0 & 0 & 3.53e-12 & 1e-12 \\
2FHLJ0428.7$-$3755 & 2.395 & 0.054 & 0.108 & 0.028 & 1.8218e-10 & 7.68e-12 \\
2FHLJ0433.6$+$2907 & 2.134 & 0.065 & 0.046 & 0.023 & 2.759e-11 & 1.71e-12 \\

%% file: Tables/fit_p2.tex
2FHLJ0440.3$-$2458 & 1.439 & 0.029 & 0.103 & 0.027 & 3.77e-12 & 1.5e-13 \\
2FHLJ0449.4$-$4349 & 1.851 & 0.056 & 0.114 & 0.044 & 1.4149e-10 & 7.48e-12 \\
2FHLJ0456.9$-$2323 & 2.453 & 0.028 & 0.051 & 0.007 & 1.1769e-10 & 3.07e-12 \\
2FHLJ0507.9$+$6737 & 1.547 & 0.064 & 0 & 0 & 2.74e-11 & 2.24e-12 \\
2FHLJ0538.6$-$4406 & 2.303 & 0.011 & 0.061 & 0.006 & 1.5054e-10 & 1.33e-12 \\
2FHLJ0543.9$-$5533 & 1.83 & 0.032 & 0.02 & 0.022 & 2.08e-11 & 1.16e-12 \\
2FHLJ0622.4$-$2604 & 1.902 & 0.088 & 0 & 0 & 2.169e-11 & 2.72e-12 \\
2FHLJ0631.0$-$2406 & 1.799 & 0.034 & 0.004 & 0.013 & 5.53e-11 & 2e-12 \\
2FHLJ0648.6$+$1516 & 1.685 & 0.016 & 0.087 & 0.01 & 2.165e-11 & 5.5e-13 \\
2FHLJ0650.7$+$2502 & 1.767 & 0.016 & 0.04 & 0.011 & 4.853e-11 & 1.33e-12 \\
2FHLJ0710.5$+$5908 & 1.688 & 0.066 & 0.007 & 0.036 & 6.71e-12 & 8.1e-13 \\
2FHLJ0721.9$+$7121 & 2.261 & 0.006 & 0.083 & 0.003 & 1.8398e-10 & 1.15e-12 \\
2FHLJ0744.2$+$7435 & 1.86 & 0.057 & 0 & 0 & 5.85e-12 & 5.9e-13 \\
2FHLJ0756.8$+$0955 & 2.297 & 0.073 & 0 & 0 & 1.071e-11 & 1.25e-12 \\
2FHLJ0805.8$+$7534 & 1.857 & 0.03 & 0.054 & 0.019 & 1.588e-11 & 8.5e-13 \\
2FHLJ0809.7$+$5218 & 1.959 & 0.02 & 0.06 & 0.012 & 4.984e-11 & 1.64e-12 \\
2FHLJ0811.6$+$0146 & 2.211 & 0.073 & 0.035 & 0.022 & 2.885e-11 & 2.03e-12 \\
2FHLJ0825.9$-$2230 & 1.968 & 0.026 & 0 & 0 & 4.778e-11 & 2.67e-12 \\
2FHLJ0847.2$+$1133 & 1.674 & 0.152 & 0 & 0 & 5.46e-12 & 1.01e-12 \\
2FHLJ0950.2$+$4553 & 1.899 & 0.15 & 0 & 0 & 4.34e-12 & 9.5e-13 \\
2FHLJ0952.2$+$7503 & 1.471 & 0.107 & 0.008 & 0.0 & 1.76e-12 & 2.6e-13 \\
2FHLJ0957.6$+$5523 & 2.221 & 0.034 & 0.068 & 0.01 & 8.894e-11 & 2.91e-12 \\
2FHLJ0958.3$+$6535 & 2.541 & 0.1 & 0.138 & 0.05 & 2.227e-11 & 1.62e-12 \\
2FHLJ1010.2$-$3119 & 1.958 & 0.098 & 0 & 0 & 1.007e-11 & 8.4e-13 \\

%% file: Tables/fit_p3.tex
2FHLJ1015.0$+$4926 & 1.917 & 0.028 & 0.046 & 0.018 & 1.0386e-10 & 4.41e-12 \\
2FHLJ1031.2$+$7441 & 2.271 & 0.092 & 0 & 0 & 4.97e-12 & 7.1e-13 \\
2FHLJ1031.4$+$5052 & 1.773 & 0.07 & 0.043 & 0.033 & 1.314e-11 & 1.31e-12 \\
2FHLJ1053.5$+$4930 & 2.047 & 0.105 & 0 & 0 & 3.86e-12 & 8.1e-13 \\
2FHLJ1058.5$+$5625 & 2.02 & 0.048 & 0.03 & 0.016 & 3.231e-11 & 1.89e-12 \\
2FHLJ1104.0$-$2331 & 1.626 & 0.088 & 0 & 0 & 6.09e-12 & 6.7e-13 \\
2FHLJ1116.9$+$2014 & 1.92 & 0.059 & 0 & 0 & 1.455e-11 & 1.33e-12 \\
2FHLJ1120.8$+$4212 & 1.583 & 0.035 & 0.046 & 0.02 & 2.047e-11 & 1.3e-12 \\
2FHLJ1125.6$-$3555 & 2.063 & 0.079 & 0 & 0 & 4.98e-12 & 7.3e-13 \\
2FHLJ1136.8$+$6737 & 1.681 & 0.042 & 0.061 & 0.024 & 7.6e-12 & 5.7e-13 \\
2FHLJ1137.9$-$1710 & 1.866 & 0.181 & 0.249 & 0.133 & 4.87e-12 & 7.8e-13 \\
2FHLJ1217.9$+$3006 & 2.077 & 0.028 & 0.063 & 0.016 & 9.429e-11 & 3.8e-12 \\
2FHLJ1221.3$+$3009 & 1.756 & 0.044 & 0.029 & 0.021 & 4.744e-11 & 3.36e-12 \\
2FHLJ1224.4$+$2435 & 1.915 & 0.063 & 0.045 & 0.026 & 1.628e-11 & 1.23e-12 \\
2FHLJ1224.7$+$2124 & 2.499 & 0.047 & 0.033 & 0.012 & 9.802e-11 & 4.45e-12 \\
2FHLJ1231.7$+$2848 & 2.166 & 0.081 & 0.08 & 0.028 & 2.041e-11 & 1.96e-12 \\
2FHLJ1256.2$-$0548 & 2.752 & 0.055 & 0.116 & 0.026 & 1.0976e-10 & 3.82e-12 \\
2FHLJ1309.5$+$4305 & 1.997 & 0.015 & 0.064 & 0.009 & 1.95e-11 & 2.8e-13 \\
2FHLJ1312.6$+$4828 & 2.254 & 0.039 & 0.024 & 0.011 & 1.835e-11 & 6.7e-13 \\
2FHLJ1404.9$+$6555 & 1.962 & 0.089 & 0.061 & 0.057 & 5.27e-12 & 7.3e-13 \\
2FHLJ1418.0$+$2543 & 1.847 & 0.071 & 0 & 0 & 3.46e-12 & 3.3e-13 \\
2FHLJ1427.3$-$4204 & 2.454 & 0.032 & 0.066 & 0.008 & 2.3028e-10 & 6.76e-12 \\
2FHLJ1428.5$+$4239 & 1.551 & 0.018 & 0.008 & 0.013 & 9.54e-12 & 3.3e-13 \\
2FHLJ1437.0$+$5639 & 1.812 & 0.053 & 0.003 & 0.024 & 7.53e-12 & 6.3e-13 \\

%% file: Tables/fit_p4.tex
2FHLJ1442.9$+$1159 & 2.202 & 0.085 & 0 & 0 & 7.58e-12 & 5.5e-13 \\
2FHLJ1506.8$+$0813 & 1.823 & 0.087 & 0.075 & 0.046 & 1.237e-11 & 1.63e-12 \\
2FHLJ1512.7$-$0906 & 2.609 & 0.007 & 0.044 & 0.002 & 1.7733e-10 & 1.14e-12 \\
2FHLJ1517.7$+$6526 & 1.708 & 0.008 & 0.047 & 0.006 & 1.29e-11 & 1.3e-13 \\
2FHLJ1548.7$-$2249 & 1.917 & 0.025 & 0 & 0 & 1.485e-11 & 5.4e-13 \\
2FHLJ1748.7$+$7004 & 2.131 & 0.034 & 0.055 & 0.02 & 4.594e-11 & 1.46e-12 \\
2FHLJ1801.2$+$7828 & 2.368 & 0.039 & 0.066 & 0.021 & 3.7e-11 & 1.12e-12 \\
2FHLJ1917.7$-$1921 & 1.928 & 0.01 & 0.026 & 0.007 & 3.824e-11 & 5.9e-13 \\
2FHLJ1936.9$-$4721 & 1.713 & 0.057 & 0 & 0 & 1.024e-11 & 8.9e-13 \\
2FHLJ1958.3$-$3011 & 1.819 & 0.068 & 0.081 & 0.048 & 1.373e-11 & 1.4e-12 \\
2FHLJ2000.9$-$1749 & 2.18 & 0.011 & 0 & 0 & 1.134e-11 & 1.9e-13 \\
2FHLJ2016.5$-$0904 & 2.044 & 0.109 & 0.067 & 0.043 & 1.295e-11 & 1.4e-12 \\
2FHLJ2116.1$+$3339 & 1.848 & 0.018 & 0.043 & 0.006 & 4.337e-11 & 5.5e-13 \\
2FHLJ2131.4$-$0914 & 1.925 & 0.069 & 0.015 & 0.046 & 7.78e-12 & 9.4e-13 \\
2FHLJ2150.3$-$1411 & 1.808 & 0.035 & 0.154 & 0.025 & 7.08e-12 & 3.5e-13 \\
2FHLJ2153.1$-$0041 & 2.04 & 0.073 & 0 & 0 & 2.05e-12 & 1.9e-13 \\
2FHLJ2158.8$-$3013 & 1.915 & 0.018 & 0.041 & 0.012 & 2.4024e-10 & 6.69e-12 \\
2FHLJ2249.9$+$3826 & 1.771 & 0.072 & 0 & 0 & 1.138e-11 & 1.1e-12 \\
2FHLJ2254.0$+$1613 & 3.111 & 0.029 & 0.201 & 0.013 & 3.0142e-10 & 5.1e-12 \\
2FHLJ2314.0$+$1445 & 1.957 & 0.098 & 0.079 & 0.047 & 7.93e-12 & 1.08e-12 \\
2FHLJ2324.7$-$4041 & 1.517 & 0.079 & 0.268 & 0.059 & 1.861e-11 & 1.26e-12 \\
2FHLJ2329.2$+$3754 & 1.838 & 0.093 & 0.174 & 0.063 & 1.065e-11 & 1.38e-12 \\
2FHLJ2340.8$+$8014 & 2.108 & 0.051 & 0.089 & 0.031 & 2.469e-11 & 1.6e-12 \\
2FHLJ2343.5$+$3438 & 1.826 & 0.056 & 0 & 0 & 5.33e-12 & 3.8e-13 \\

%% file: jcap_ver.bbl
\begin{thebibliography}{10}

\bibitem{ALPS:btsm1}
S.~Chang, S.~Tazawa, and M.~Yamaguchi, ``{Axion model in extra dimensions with
  TeV scale gravity},'' {\em Physical Review D}, vol.~D61, p.~084005, 2000.

\bibitem{ALPS:btsm2}
N.~Turok, ``{Almost Goldstone bosons from extra dimensional gauge theories},''
  {\em Physical Review Letters}, vol.~76, pp.~1015--1018, 1996.

\bibitem{ALPS:btsm3}
P.~Svrcek and E.~Witten, ``{Axions In String Theory},'' {\em Journal of High
  Energy Physics}, vol.~06, p.~051, 2006.

\bibitem{ALPS:btsm4}
A.~Arvanitaki, S.~Dimopoulos, S.~Dubovsky, N.~Kaloper, and J.~March-Russell,
  ``{String Axiverse},'' {\em Physical Review D}, vol.~D81, p.~123530, 2010.

\bibitem{ALPS:btsm5}
C.~Coriano and N.~Irges, ``{Windows over a New Low Energy Axion},'' {\em
  Physics Letters B}, vol.~B651, pp.~298--305, 2007.

\bibitem{ALPS:btsm6}
H.~Baer, M.~Haider, S.~Kraml, S.~Sekmen, and H.~Summy, ``{Cosmological
  consequences of Yukawa-unified SUSY with mixed axion/axino cold and warm dark
  matter},'' {\em Journal of Cosmology and Astroparticle Physics}, vol.~0902,
  p.~002, 2009.

\bibitem{AX:PQ}
R.~D. {Peccei}, ``{QCD, strong CP and axions.},'' {\em Journal of Korean
  Physical Society}, vol.~29, pp.~S199--S208, Sept. 1996.

\bibitem{cor7}
F.~Wilczek, ``Problem of strong $p$ and $t$ invariance in the presence of
  instantons,'' {\em Physical Review Letters}, vol.~40, pp.~279--282, Jan 1978.

\bibitem{cor8}
S.~Weinberg, ``A new light boson?,'' {\em Physical Review Letters}, vol.~40,
  pp.~223--226, Jan 1978.

\bibitem{2014arXiv1407.0546R;AXIONS_OVERVIEW_RINGWALD}
A.~{Ringwald}, ``{Axions and Axion-Like Particles},'' {\em ArXiv e-prints},
  p.~arXiv:1407.0546, July 2014.

\bibitem{PRESKILL1983127}
J.~Preskill, M.~B. Wise, and F.~Wilczek, ``Cosmology of the invisible axion,''
  {\em Physics Letters B}, vol.~120, no.~1, pp.~127 -- 132, 1983.

\bibitem{11n2:dine}
M.~{Dine} and W.~{Fischler}, ``{The not-so-harmless axion},'' {\em Physics
  Letters B}, vol.~120, pp.~137--141, Jan. 1983.

\bibitem{11n3:abbott}
L.~Abbott and P.~Sikivie, ``{A Cosmological Bound on the Invisible Axion},''
  {\em Physics Letters B}, vol.~120, pp.~133--136, 1983.

\bibitem{Sikivie:2009fv}
P.~Sikivie, ``{Dark matter axions},'' {\em International Journal of Modern
  Physics}, vol.~A25, pp.~554--563, 2010.

\bibitem{EBL:bitteau}
J.~{Biteau} and D.~A. {Williams}, ``{The Extragalactic Background Light, the
  Hubble Constant, and Anomalies: Conclusions from 20 Years of TeV Gamma-ray
  Observations},'' {\em "The Astrophysical Journal"}, vol.~812, p.~60, Oct.
  2015.

\bibitem{EBL:dwek}
E.~{Dwek} and F.~{Krennrich}, ``{The extragalactic background light and the
  gamma-ray opacity of the universe},'' {\em Astroparticle Physics}, vol.~43,
  pp.~112--133, Mar. 2013.

\bibitem{FIELDS:faraday}
P.~P. {Kronberg} and J.~J. {Perry}, ``{Absorption lines, Faraday rotation, and
  magnetic field estimates for QSO absorption-line clouds},'' {\em "The
  Astrophysical Journal"}, vol.~263, pp.~518--532, Dec. 1982.

\bibitem{IGM:ns_cmb4}
T.~{Kahniashvili}, Y.~{Maravin}, and A.~{Kosowsky}, ``{Faraday rotation limits
  on a primordial magnetic field from Wilkinson Microwave Anisotropy Probe
  five-year data},'' {\em Physical Review D}, vol.~80, p.~023009, July 2009.

\bibitem{IGM:ns_or1}
P.~P. {Kronberg} and M.~{Simard-Normandin}, ``{New evidence on the origin of
  rotation measures in extragalactic radio sources},'' {\em Nature}, vol.~263,
  pp.~653--656, Oct. 1976.

\bibitem{PAG:IC_obs}
M.~{Murgia}, F.~{Govoni}, L.~{Feretti}, G.~{Giovannini}, D.~{Dallacasa},
  R.~{Fanti}, G.~B. {Taylor}, and K.~{Dolag}, ``{Magnetic fields and Faraday
  rotation in clusters of galaxies},'' {\em Astronomy and Astrophysics},
  vol.~424, pp.~429--446, Sept. 2004.

\bibitem{bfields}
L.~M. {Widrow}, ``{Origin of galactic and extragalactic magnetic fields},''
  {\em Reviews of Modern Physics}, vol.~74, pp.~775--823, Jan 2002.

\bibitem{FIELDS:rev}
J.~Han, ``Observing interstellar and intergalactic magnetic fields,'' {\em
  Annual Review of Astronomy and Astrophysics}, vol.~55, no.~1, pp.~111--157,
  2017.

\bibitem{TheFermi-LAT:2016zue}
M.~Ajello {\em et~al.}, ``{Search for Spectral Irregularities due to
  Photon-Axionlike-Particle Oscillations with the Fermi Large Area
  Telescope},'' {\em Physical Review Letters}, vol.~116, no.~16, p.~161101,
  2016.

\bibitem{hooper2007}
D.~{Hooper} and P.~D. {Serpico}, ``{Detecting Axionlike Particles with Gamma
  Ray Telescopes},'' {\em Physical Review Letters}, vol.~99, p.~231102, Dec
  2007.

\bibitem{hardening1}
A.~{Dom{\'\i}nguez}, M.~A. {S{\'a}nchez-Conde}, and F.~{Prada}, ``{Axion-like
  particle imprint in cosmological very-high-energy sources},'' {\em Journal of
  Cosmology and Astro-Particle Physics}, vol.~2011, p.~020, Nov. 2011.

\bibitem{hardening2}
D.~{Horns}, L.~{Maccione}, M.~{Meyer}, A.~{Mirizzi}, D.~{Montanino}, and
  M.~{Roncadelli}, ``{Hardening of TeV gamma spectrum of active galactic nuclei
  in galaxy clusters by conversions of photons into axionlike particles},''
  {\em Physical Review D}, vol.~86, p.~075024, Oct. 2012.

\bibitem{hardening3}
D.~{Montanino}, F.~{Vazza}, A.~{Mirizzi}, and M.~{Viel}, ``{Enhancing the
  Spectral Hardening of Cosmic TeV Photons by Mixing with Axionlike Particles
  in the Magnetized Cosmic Web},'' {\em Physical Review Letters}, vol.~119,
  p.~101101, Sept. 2017.

\bibitem{paper_igmf}
A.~{de Angelis}, M.~{Roncadelli}, and O.~{Mansutti}, ``{Evidence for a new
  light spin-zero boson from cosmological gamma-ray propagation?},'' {\em
  Physical Review D}, vol.~76, p.~121301, Dec. 2007.

\bibitem{hardening4}
M.~A. {S{\'a}nchez-Conde}, D.~{Paneque}, E.~{Bloom}, F.~{Prada}, and
  A.~{Dom{\'{\i}}nguez}, ``{Hints of the existence of axionlike particles from
  the gamma-ray spectra of cosmological sources},'' {\em Physical Review D},
  vol.~79, p.~123511, June 2009.

\bibitem{hess_limits}
A.~{Abramowski} {\em et~al.}, ``{Constraints on axionlike particles with
  H.E.S.S. from the irregularity of the PKS 2155-304 energy spectrum},'' {\em
  Physical Review D}, vol.~88, p.~102003, Nov. 2013.

\bibitem{2009ApJ...697.1071A:fermilat_atwood2009}
W.~B. {Atwood} {\em et~al.}, ``{The Large Area Telescope on the Fermi Gamma-Ray
  Space Telescope Mission},'' {\em "The Astrophysical Journal"}, vol.~697,
  pp.~1071--1102, June 2009.

\bibitem{PAG:agn}
F.~{Tavecchio}, M.~{Roncadelli}, and G.~{Galanti}, ``{Photons to axion-like
  particles conversion in Active Galactic Nuclei},'' {\em Physics Letters B},
  vol.~744, pp.~375--379, May 2015.

\bibitem{2017A&A...603A..34F:EBL_FRV_current}
A.~{Franceschini} and G.~{Rodighiero}, ``{The extragalactic background light
  revisited and the cosmic photon- photon opacity},'' {\em Astronomy and
  Astrophysics}, vol.~603, p.~A34, July 2017.

\bibitem{pair_production}
R.~J. Gould and G.~P. Schr\'eder, ``Pair production in photon-photon
  collisions,'' {\em Physical Review}, vol.~155, pp.~1404--1407, Mar 1967.

\bibitem{cor9}
A.~Nikishov, ``Absorption of high energy photons in the universe,'' {\em
  Journal of Experimental and Theoretical Physics}, vol.~Vol: 41, 08 1961.

\bibitem{cor3}
A.~{Dom{\'{\i}}nguez} and F.~{Prada}, ``{Measurement of the Expansion Rate of
  the Universe from {$\gamma$}-Ray Attenuation},'' {\em Astrophysical Journal,
  Letters}, vol.~771, p.~L34, July 2013.

\bibitem{EBL:sfh2}
{Fermi-LAT Collaboration}, S.~{Abdollahi}, {\em et~al.}, ``{A gamma-ray
  determination of the Universe's star formation history},'' {\em Science},
  vol.~362, pp.~1031--1034, Nov. 2018.

\bibitem{pubv2}
A.~{Dom{\'\i}nguez}, R.~{Wojtak}, J.~{Finke}, M.~{Ajello}, K.~{Helgason},
  F.~{Prada}, A.~{Desai}, V.~{Paliya}, L.~{Marcotulli}, and D.~H. {Hartmann},
  ``{A New Measurement of the Hubble Constant and Matter Content of the
  Universe Using Extragalactic Background Light {\ensuremath{\gamma}}-Ray
  Attenuation},'' {\em The Astrophysical Journal}, vol.~885, p.~137, Nov. 2019.

\bibitem{EBL:challenges}
M.~G. {Hauser} and E.~{Dwek}, ``{The Cosmic Infrared Background: Measurements
  and Implications},'' {\em Annual Review of Astronomy and Astrophysics},
  vol.~39, pp.~249--307, Jan. 2001.

\bibitem{EBL:latest}
{H.~E.~S.~S. Collaboration}, ``{Measurement of the EBL spectral energy
  distribution using the VHE {\ensuremath{\gamma}}-ray spectra of H.E.S.S.
  blazars},'' {\em Astronomy and Astrophysics}, vol.~606, p.~A59, Oct. 2017.

\bibitem{EBL:magic1}
V.~A. {Acciari} {\em et~al.}, ``{Measurement of the extragalactic background
  light using MAGIC and Fermi-LAT gamma-ray observations of blazars up to z =
  1},'' {\em Monthly Notices of the Royal Astronomical Society}, vol.~486,
  pp.~4233--4251, July 2019.

\bibitem{EBL:veritas1}
A.~U. {Abeysekara} {\em et~al.}, ``{Measurement of the Extragalactic Background
  Light Spectral Energy Distribution with VERITAS},'' {\em "The Astrophysical
  Journal"}, vol.~885, p.~150, Nov. 2019.

\bibitem{EBL:lat1}
A.~A. {Abdo} {\em et~al.}, ``{Fermi Large Area Telescope Constraints on the
  Gamma-ray Opacity of the Universe},'' {\em "The Astrophysical Journal"},
  vol.~723, pp.~1082--1096, Nov. 2010.

\bibitem{cor6}
A.~{Dom{\'{\i}}nguez} and M.~{Ajello}, ``{Spectral Analysis of Fermi-LAT
  Blazars above 50 GeV},'' {\em Astrophysical Journal, Letters}, vol.~813,
  p.~L34, Nov. 2015.

\bibitem{EBL:dominguez}
A.~{Dom{\'\i}nguez}, J.~R. {Primack}, D.~J. {Rosario}, F.~{Prada}, R.~C.
  {Gilmore}, S.~M. {Faber}, D.~C. {Koo}, R.~S. {Somerville}, M.~A.
  {P{\'e}rez-Torres}, P.~{P{\'e}rez-Gonz{\'a}lez}, J.~S. {Huang}, M.~{Davis},
  P.~{Guhathakurta}, P.~{Barmby}, C.~J. {Conselice}, M.~{Lozano}, J.~A.
  {Newman}, and M.~C. {Cooper}, ``{Extragalactic background light inferred from
  AEGIS galaxy-SED-type fractions},'' {\em Monthly Notices of the RAS},
  vol.~410, pp.~2556--2578, Feb. 2011.

\bibitem{EBL:forward1}
J.~D. {Finke}, S.~{Razzaque}, and C.~D. {Dermer}, ``{Modeling the Extragalactic
  Background Light from Stars and Dust},'' {\em "The Astrophysical Journal"},
  vol.~712, pp.~238--249, Mar. 2010.

\bibitem{cor4}
G.~G. {Fazio} and F.~W. {Stecker}, ``{Predicted High Energy Break in the
  Isotropic Gamma Ray Spectrum: a Test of Cosmological Origin},'' {\em Nature},
  vol.~226, pp.~135--136, Apr. 1970.

\bibitem{cor5}
A.~{Dom{\'{\i}}nguez}, J.~D. {Finke}, F.~{Prada}, J.~R. {Primack}, F.~S.
  {Kitaura}, B.~{Siana}, and D.~{Paneque}, ``{Detection of the Cosmic
  {$\gamma$}-Ray Horizon from Multiwavelength Observations of Blazars},'' {\em
  "The Astrophysical Journal"}, vol.~770, p.~77, June 2013.

\bibitem{2FHL:catalog}
M.~{Ackermann} {\em et~al.}, ``{2FHL: The Second Catalog of Hard Fermi-LAT
  Sources},'' {\em The Astrophysical Journal Supplement Series}, vol.~222,
  p.~5, Jan. 2016.

\bibitem{ALPS:oscillations}
G.~Raffelt and L.~Stodolsky, ``{Mixing of the Photon with Low Mass
  Particles},'' {\em Physical Review D}, vol.~D37, p.~1237, 1988.

\bibitem{IGM:ns}
A.~{Neronov} and D.~V. {Semikoz}, ``{Sensitivity of {\ensuremath{\gamma}}-ray
  telescopes for detection of magnetic fields in the intergalactic medium},''
  {\em Physical Review D}, vol.~80, p.~123012, Dec. 2009.

\bibitem{IGM:cosmo_book}
P.~J.~E. Peebles, {\em Principles of physical cosmology}.
\newblock Princeton, N.J. : Princeton University Press, 1993.
\newblock p. 685-709.

\bibitem{GOLD:formulas}
A.~{de Angelis}, G.~{Galanti}, and M.~{Roncadelli}, ``{Relevance of axionlike
  particles for very-high-energy astrophysics},'' {\em Physical Review D},
  vol.~84, p.~105030, Nov. 2011.

\bibitem{cor10}
C.~{Cs{\'a}ki}, N.~{Kaloper}, M.~{Peloso}, and J.~{Terning}, ``{Super-GZK
  photons from photon-axion mixing},'' {\em "Journal of Cosmology and
  Astroparticle Physics"}, vol.~5, p.~005, May 2003.

\bibitem{PAG:IC_manuel}
M.~{Meyer}, D.~{Montanino}, and J.~{Conrad}, ``{On detecting oscillations of
  gamma rays into axion-like particles in turbulent and coherent magnetic
  fields},'' {\em Journal of Cosmology and Astro-Particle Physics}, vol.~2014,
  p.~003, Sept. 2014.

\bibitem{LAT:pass8}
W.~{Atwood}, A.~{Albert}, L.~{Baldini}, M.~{Tinivella}, J.~{Bregeon},
  M.~{Pesce-Rollins}, C.~{Sgro}, P.~{Bruel}, E.~{Charles}, A.~{Drlica-Wagner},
  A.~{Franckowiak}, T.~{Jogler}, L.~{Rochester}, T.~{Usher}, M.~{Wood},
  J.~{Cohen-Tanugi}, and {S. Zimmer for the Fermi-LAT Collaboration}, ``{Pass
  8: Toward the Full Realization of the Fermi-LAT Scientific Potential},'' {\em
  arXiv e-prints}, p.~arXiv:1303.3514, Mar. 2013.

\bibitem{BLZ:spectra}
J.~P. {van den Berg}, M.~{B{\"o}ttcher}, A.~{Dom{\'\i}nguez}, and
  M.~{L{\'o}pez-Moya}, ``{Systematic Physical Characterization of the
  {\ensuremath{\gamma}}-Ray Spectra of 2FHL Blazars},'' {\em "The Astrophysical
  Journal"}, vol.~874, p.~47, Mar. 2019.

\bibitem{cowan:stats}
G.~Cowan, {\em Statistical data analysis}.
\newblock Oxford University Press, USA, 1998.

\bibitem{AP:supernovae}
A.~{Payez}, C.~{Evoli}, T.~{Fischer}, M.~{Giannotti}, A.~{Mirizzi}, and
  A.~{Ringwald}, ``{Revisiting the SN1987A gamma-ray limit on ultralight
  axion-like particles},'' {\em Journal of Cosmology and Astro-Particle
  Physics}, vol.~2015, p.~006, Feb. 2015.

\bibitem{alps2s}
R.~{B{\"a}hre}, B.~{D{\"o}brich}, J.~{Dreyling-Eschweiler}, S.~{Ghazaryan},
  R.~{Hodajerdi}, D.~{Horns}, F.~{Januschek}, E.~A. {Knabbe}, A.~{Lindner},
  D.~{Notz}, A.~{Ringwald}, J.~E. {von Seggern}, R.~{Stromhagen}, D.~{Trines},
  and B.~{Willke}, ``{Any light particle search II {\textemdash} Technical
  Design Report},'' {\em Journal of Instrumentation}, vol.~8, p.~T09001, Sep
  2013.

\bibitem{iaxos}
I.~G. Irastorza {\em et~al.}, ``{Future axion searches with the International
  Axion Observatory (IAXO)},'' {\em Journal of Physics: Conference Series},
  vol.~460, p.~012002, 2013.

\bibitem{alps:cdm_mm}
P.~{Arias}, D.~{Cadamuro}, M.~{Goodsell}, J.~{Jaeckel}, J.~{Redondo}, and
  A.~{Ringwald}, ``{WISPy cold dark matter},'' {\em Journal of Cosmology and
  Astro-Particle Physics}, vol.~2012, p.~013, June 2012.

\bibitem{jvla}
R.~{Perley}, P.~{Napier}, J.~{Jackson}, B.~{Butler}, B.~{Carlson}, D.~{Fort},
  P.~{Dewdney}, B.~{Clark}, R.~{Hayward}, S.~{Durand }, M.~{Revnell}, and
  M.~{McKinnon}, ``{The Expanded Very Large Array},'' {\em IEEE Proceedings},
  vol.~97, pp.~1448--1462, Aug 2009.

\bibitem{alma}
A.~{Wootten} and A.~R. {Thompson}, ``{The Atacama Large
  Millimeter/Submillimeter Array},'' {\em IEEE Proceedings}, vol.~97,
  pp.~1463--1471, Aug 2009.

\bibitem{ska}
B.~M. {Gaensler}, R.~{Beck}, and L.~{Feretti}, ``{The origin and evolution of
  cosmic magnetism},'' {\em New Astronomy Review}, vol.~48, pp.~1003--1012, Dec
  2004.

\bibitem{desai19}
A.~{Desai}, K.~{Helgason}, M.~{Ajello}, V.~{Paliya}, A.~{Dom{\'\i}nguez},
  J.~{Finke}, and D.~{Hartmann}, ``{A GeV-TeV Measurement of the Extragalactic
  Background Light},'' {\em "The Astrophysical Journal"}, vol.~874, p.~L7, Mar
  2019.

\bibitem{cta_2016}
C.~{Bigongiari} and {CTA Consortium}, ``{The Cherenkov Telescope Array},'' {\em
  Nuclear and Particle Physics Proceedings}, vol.~279, pp.~174--181, Oct 2016.

\bibitem{ind:ciber}
K.~{Kohri} and H.~{Kodama}, ``{Axion-like particles and recent observations of
  the cosmic infrared background radiation},'' {\em Physical Review D},
  vol.~96, p.~051701, Sept. 2017.

\bibitem{ind:gamma}
G.~{Galanti}, M.~{Roncadelli}, A.~{De Angelis}, and G.~F. {Bignami}, ``{Hint at
  an axion-like particle from the redshift dependence of blazar spectra},''
  {\em Monthly Notices of the Royal Astronomical Society}, vol.~493,
  pp.~1553--1564, Apr. 2020.

\bibitem{oscillations:raffelt}
G.~Raffelt and L.~Stodolsky, ``{Mixing of the Photon with Low Mass
  Particles},'' {\em Physical Review D}, vol.~D37, p.~1237, 1988.

\bibitem{HSW:lag}
W.~Heisenberg and H.~Euler, ``{Folgerungen aus der Diracschen Theorie des
  Positrons},'' {\em Zeitschrift f{\"u}r Physik}, vol.~98, pp.~714--732, 1936.

\end{thebibliography}
